\documentclass[aps,prfluids,preprint,superscriptaddress,onecolumn]{revtex4-2}

\usepackage{graphicx}%
\usepackage{amsmath,amssymb,amsfonts}%
\usepackage{amsthm}%
\usepackage{mathrsfs}%
\usepackage{xcolor}
\definecolor{color1}{HTML}{303a2b}
\definecolor{color2}{HTML}{00798c}
\definecolor{color3}{HTML}{d1495b}
\definecolor{color4}{HTML}{eddb53}
\usepackage[separate-uncertainty=true]{siunitx}  
\usepackage{mathrsfs}
\usepackage{dcolumn}

\usepackage{caption}
\usepackage{subcaption}

\usepackage{hyperref}

\usepackage{color}

\newcommand{\EOA}{\text{EOA}}
\newcommand{\GOA}{\text{GOA}}
\newcommand{\BSA}{\text{BSA}}

\newcommand{\WSS}{\text{WSS}} 
\newcommand{\aAO}{\text{aAO}} 
\newcommand{\aWSS}{\overline{\WSS}} 
\newcommand{\TPD}{\text{TPD}} 
\newcommand{\vel}{\mathbf{u}} 
\newcommand{\pos}{\mathbf{x}} 
\newcommand{\viscstress}{\boldsymbol{\tau}} 
\newcommand{\taueq}{\tau_\text{eq}} 
\newcommand{\hemolL}{\text{H}_L} 
\newcommand{\HI}{\text{HI}} 
\newcommand{\deftens}{\mathbf{S}} 
\newcommand{\rotrate}{\boldsymbol{\Omega}} 
\newcommand{\straintens}{\mathbf{E}} 
\newcommand{\Id}{\mathbf{I}} 
\newcommand{\vorticitytens}{\mathbf{W}} 
\newcommand{\DI}{D_\text{I}} 
\newcommand{\DE}{D_\text{E}} 

\newcommand{\AD}{\text{AD}} 
\DeclareSIUnit\dyne{dyne}

\begin{document}


\title{Hemodynamic effects of intra- and supra- deployment locations for a bio-prosthetic aortic valve}


\author{Martino Andrea \surname{Scarpolini}}
\affiliation{Gran Sasso Science Institute (GSSI), L'Aquila, Italy.}
\affiliation{INFN Tor Vergata, Rome, Italy.}
\author{Giovanni Vagnoli}
\affiliation{Gran Sasso Science Institute (GSSI), L'Aquila, Italy.}
\author{Fabio Guglietta}
\affiliation{Tor Vergata University of Rome, Rome, Italy.}
\affiliation{INFN Tor Vergata, Rome, Italy.}
\author{Roberto Verzicco}
\affiliation{Gran Sasso Science Institute (GSSI), L'Aquila, Italy.}
\affiliation{Tor Vergata University of Rome, Rome, Italy.}
\author{Francesco Viola}
\affiliation{Gran Sasso Science Institute (GSSI), L'Aquila, Italy.}
\email[]{francesco.viola@gssi.it}


\date{\today}

\begin{abstract}
Aortic valve replacement is a key surgical procedure for treating aortic valve pathologies, such as stenosis and regurgitation. The precise placement of the prosthetic valve relative to the native aortic annulus plays a critical role in the post-operative hemodynamics. This study investigates how the positioning of a biological prosthetic valve--either intra-annular (within the native annulus) or supra-annular (slightly downstream, in the widened portion of the aortic root)--affects cardiac fluid dynamics. Using high-fidelity numerical simulations on a patient-specific left heart model derived from CT imaging, we simulate physiological flow conditions to isolate the impact of valve placement. Unlike previous clinical studies that compare different patients and valve models, our approach evaluates the same valve in both positions within a single virtual patient, ensuring a controlled comparison. Key hemodynamic parameters are assessed, including transvalvular pressure drop, effective orifice area, wall shear stress, and hemolysis. Results reveal that supra-annular implantation offers significant advantages: lower pressure gradients, larger orifice area, and reduced shear-induced stress. Furthermore, hemolysis analysis using advanced red blood cell stress models indicates a decreased risk of blood damage in the supra-annular configuration. These findings offer valuable insights to guide valve selection and implantation strategies, ultimately supporting improved patient outcomes.

\end{abstract}


\maketitle

\section{Introduction}
\label{sec:intro}
\begin{figure}[h]
  \centering
  \includegraphics[width=0.47\textwidth]{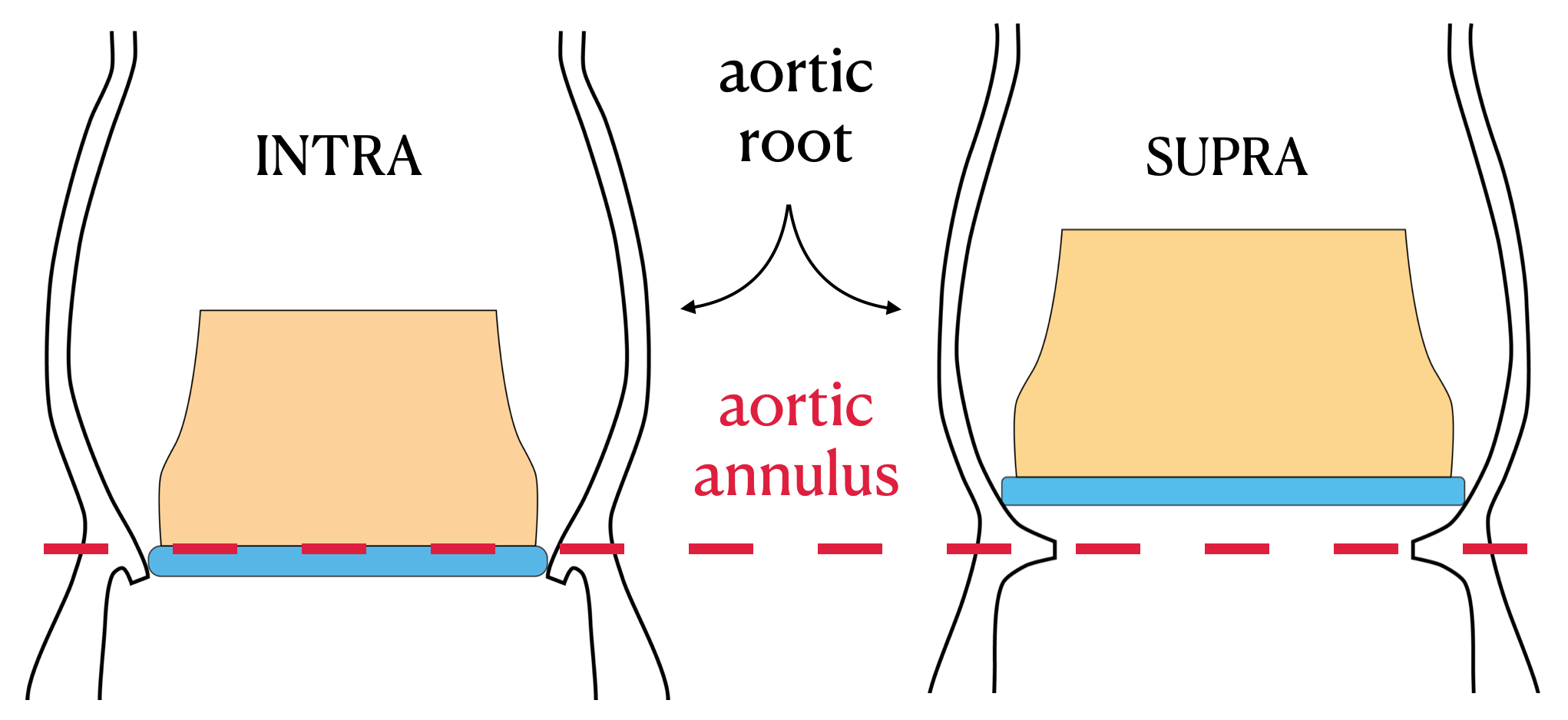}
  \caption{Intra and Supra-annular positioning for prothesic valves during aortic valve replacement interventions.}
  \label{fig:Figure1}
\end{figure}

The aortic valve plays a crucial role in maintaining unidirectional blood flow from the left ventricle to the aorta. However, anatomical or functional abnormalities can lead to two major valvular diseases: aortic stenosis (AS) and aortic regurgitation (AR). Both conditions significantly impact cardiac function and often necessitate surgical or transcatheter interventions. AR occurs when the valve fails to close properly, allowing blood to flow back into the left ventricle during diastole.
AS is characterized by a progressive narrowing of the valve orifice, increasing resistance to left ventricular outflow, and it is the most common valve lesion requiring surgical or transcatheter intervention, with its incidence in Europe and North America rising due to the aging of the population \cite{vahanian2021_ESC_guidelines_VHD}. The primary cause of AS in these regions is calcific degeneration, although congenital abnormalities, such as bicuspid aortic valve, can also contribute.\\
\indent Echocardiography, particularly Doppler echocardiography, is the gold standard for diagnosing valve stenosis and assessing its severity \cite{baumgartner2017ESC_AVS_assessment_echocardiography}. It provides crucial information on valve anatomy, leaflets mobility, and calcification while enabling the measurement of key hemodynamic parameters, including: transvalvular pressure drop, peak transvalvular velocity and valve orifice area. These indicators, along with their established diagnostic thresholds, guide clinical decision-making and determine whether surgical intervention is required.\\
\indent When surgical repair is not feasible, valve replacement is the standard treatment for severe AS and AR.
Two main types of prosthetic valves exist: mechanical and biological. The choice depends on factors such as patient life expectancy, lifestyle, comorbidities, and anticoagulation requirements.
\indent In addition to the prothesic valve model, other factors significantly influence the outcome of the surgery, such as the orientation of the prosthesis relative to the native annular axis. Studies have shown that even slight misalignments can impact postoperative hemodynamics, potentially increasing the risk of leaflet thrombosis by altering blood flow patterns and affecting sinus washout efficiency \cite{hatoum2019AOV_tilted_deployment}. This suggests that proper positioning and alignment of the valve are critical to optimizing long-term function and minimizing complications.\\
\indent An important consideration to take into account is also the size of the aortic annulus (and of the prosthesis, consequently), which can be smaller than normal in many patients with aortic valve disease. In these cases, prosthesis-patient mismatch (PPM) can occur, i.e. when the effective orifice area of the prosthesis is too small relative to the patient’s body size. This anatomical constraint poses challenges during prosthesis implantation and can impact postoperative hemodynamic performance. To address this issue, prosthetic valve deployment becomes a critical factor in optimizing surgical outcomes. One approach is annular enlargement, where surgeons expand the native annulus to accommodate a larger prosthesis. However, this technique increases surgical complexity and operative risk \cite{zhang2010intra-supra-literature-review-very-old}. An alternative strategy is supra-annular valve positioning, where the prosthesis is implanted slightly above the native intra-annular position, within the aortic root near the Valsalva sinuses. As shown in Figure~\ref{fig:Figure1}, this region provides a naturally larger diameter, allowing for a larger effective orifice area \cite{zhang2010intra-supra-literature-review-very-old}.\\
\indent Clinical in-vivo studies have shown that supra-annular valve generally results in better hemodynamic performance compared to intra-annular deployment \cite{okuno2019PPM_supra_intra, kim2019supra-intra-clinical-outcome-comparison, takahashi2023_intra_supra_hemo_difference_MRI}. This configuration has been associated with lower transvalvular pressure gradients, larger orifice areas, and potentially better long-term outcomes, such as left ventricular mass index regression. However, longitudinal follow-up data beyond five years remain limited, making long-term hemodynamic differences between the two strategies unclear. \citet{takahashi2023_intra_supra_hemo_difference_MRI}, for example, investigated early hemodynamic effects (approximately six days post-procedure) in 98 patients who underwent transcatheter aortic valve replacement using 4D flow cardiac magnetic resonance imaging (MRI) \cite{takahashi2023_intra_supra_hemo_difference_MRI}. Their findings suggest that the supra-annular configuration might provide superior performance in terms of orifice area, mean pressure gradient, and wall shear stress.\\
\indent Despite these findings, clinical studies face inherent limitations. Imaging techniques such as echocardiography provide valuable anatomical information but lack the resolution to capture fine structural details, particularly for rapidly moving leaflets or stenotic valves. MRI can acquire three-dimensional blood velocity distributions, but is constrained by limited resolution and time-averaged data, preventing detailed measurements of velocity fluctuations, wall shear stresses, and pressure distributions. Additionally, clinical studies must be non-invasive, meaning that direct in vivo comparisons between different implantation strategies in the same patient are not feasible. As a result, it is difficult to isolate the specific impact of increased valve size and positioning on hemodynamics.\\
\indent To address these limitations, we employ here fluid-structure interaction to investigate how prosthetic valve deployment affects hemodynamics within an anatomically detailed left heart model. Specifically, we select a small-sized heart, reproducing the scenario where patients with small aortic diameters suggest the surgeon to evaluate alternative surgical strategies. We consider a bioprosthetic valve model resembling the LivaNova Crown PRT, specifically chosen because, unlike most prosthetic valves, it can be implanted in both configurations. This unique feature allows us to perform a controlled, high-fidelity numerical comparison within the same virtual patient, ensuring identical anatomical and functional conditions and isolating the effects of deployment strategy. Furthermore, while the large scale effects of a valve with increased diameter might be partially guessed, our numerical approach enables the evaluation of hemodynamic parameters that are difficult or impossible to measure in clinical practice, such as velocity gradients and wall shear stress, which result from the dynamics of (chaotic) small scale structures. These insights contribute to a better understanding of how valve design and positioning influence long-term outcomes.\\
\indent The paper is organized as follows: section~\ref{sec:mm} presents the computational framework of our numerical study along with the anatomical and geometrical details of the left heart and bioprosthetic valve models. The governing equations and numerical solvers are then introduced, along with the nudging strategy employed to enforce a physiologically accurate left ventricular volume law. Successively, a brief review of numerical hemolysis models is provided, outlining the different approaches considered in this study. Section~\ref{sec:results} presents the results, starting with an overview of the computational setup and then an accurate comparison of the Wiggers diagram, valve orifice area metrics, transvalvular pressure drop, wall shear stress and hemolysis. Finally, Section~\ref{sec:disc} provides a quantitative discussion of the results, comparing findings with existing clinical and numerical studies from the literature.

\section{Material and Methods} \label{sec:mm}
\begin{figure*}[h]
  \centering
  \includegraphics[width=0.97\textwidth]{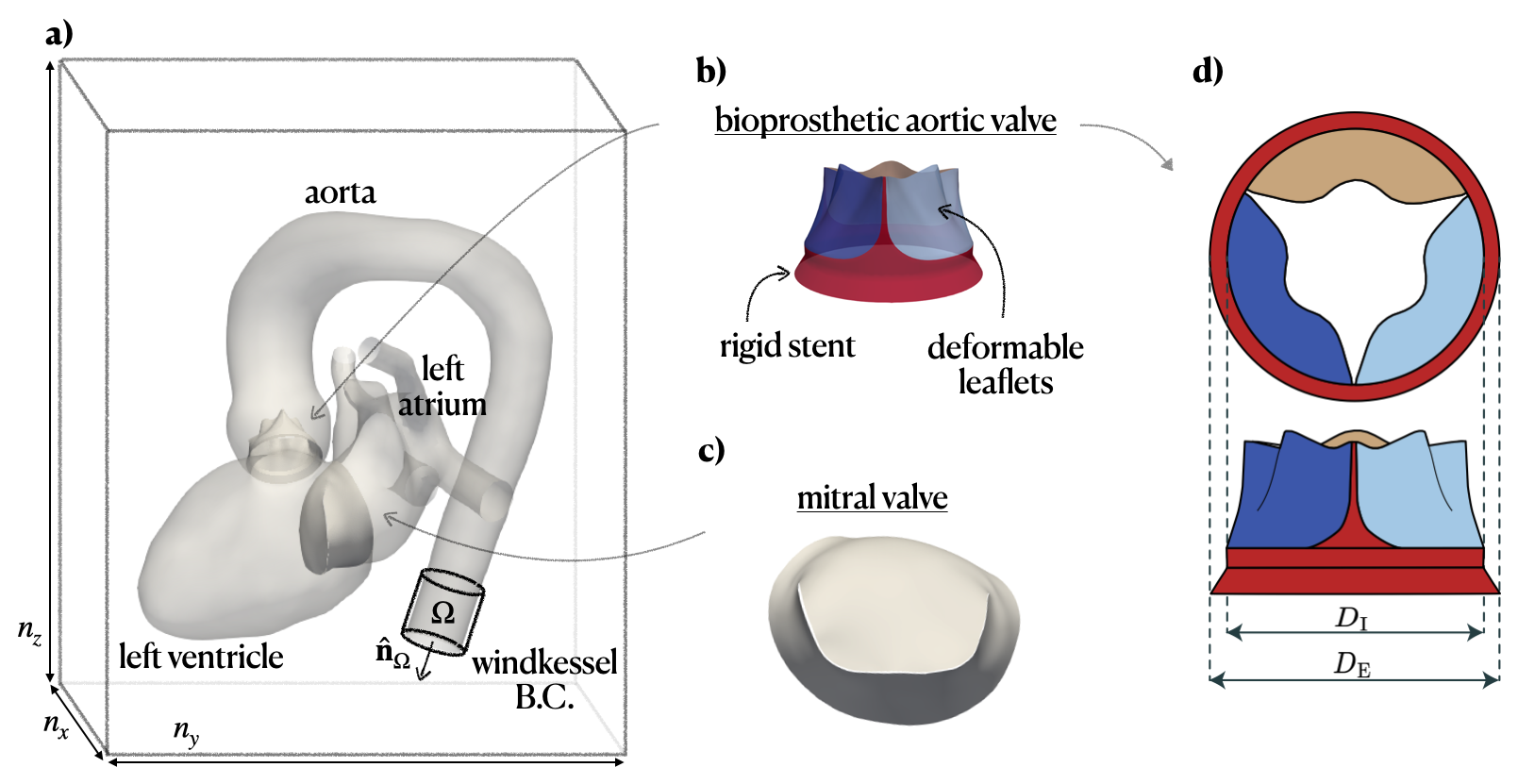}
  \caption{Panel \textbf{a)}: Setup of the computational model. The box delineates the eulerian fluid domain, where the cardiovascular structures are immersed. Panel \textbf{b)} and \textbf{c)} show the prothesic aortic valve and natural mitral valve. Panel \textbf{d)} shows geometrical details of the bioprosthetic aortic valve, with internal $\DI$ and external $\DE$ diameters specified.}
  \label{fig:Figure_setup}
\end{figure*}

Our computational setup is based on a high-fidelity geometric model of the left heart, extracted from a thoracic CT scan of a 44-year-old male patient with no diagnosed cardiovascular disease. The imaging data was retrospectively analyzed and acquired using a 320-detector row CT scanner (Aquilion One PRISM edition, Canon Medical Systems) during the diastolic phase, with a spatial resolution of $0.728\times0.728\times1.0$~\unit{\mm}. The heart model was generated through semi-automatic segmentation using 3DSlicer software \cite{3dslicer}. Figure~\ref{fig:Figure_setup}a illustrates the computational domain, where the outlined box delineates the boundaries of the Eulerian fluid domain, within which the cardiovascular structures are immersed. The left heart model includes left atrium, left ventricle, and aorta. The left atrium, with an approximate volume of 80~\unit{\ml}, includes auricle (left atrial appendage) and pulmonary veins, which serve as the entry points for blood into the heart. The mitral valve geometry (Figure~\ref{fig:Figure_setup}c) was adopted from a previous study \cite{meschini2019mitral_stenosis} and adjusted to fit the current mitral annulus, as direct segmentation of the valve from the CT scan was not feasible. The left ventricle volume is approximately 114~\unit{\ml}, while the aortic annulus diameter is 21~\unit{\mm}, corresponding to a small-sized heart. This anatomical configuration is clinically relevant, as patients with small aortic diameters may require alternative deployment positions. The aortic geometry extends through the entire thoracic region, including the ascending aorta, aortic arch, and descending aorta. The prosthesis consists of a tri-leaflet biological valve made from bovine pericardium, supported by a crown-shaped rigid stent. The computational representation of the valve is shown in Figure~\ref{fig:Figure_setup}b,d, where the red region highlights the rigid stent structure.\\
\indent To analyze the impact of supra-annular vs. intra-annular deployment, we perform and compare two numerical simulations, each corresponding to one implantation strategy. The prosthetic valve sizes were selected based on the available specifications from the medical device catalog. The rigid stent is characterized by its internal diameter ($\DI$) and external diameter ($\DE$), as illustrated in Figure~\ref{fig:Figure_setup}d. For the intra-annular deployment, the valve size is selected as to match the aortic annulus, $\DE=21$~\unit{\mm} and $\DI=18.6$~\unit{\mm}. In contrast, for the supra-annular deployment, the next available size in the catalog is used, with slightly larger dimensions: $\DE=23$~\unit{\mm} and $\DI=20.7$~\unit{\mm}. This setup allows a controlled comparison within the same anatomical model and valve design, isolating the impact of the deployment strategy on hemodynamics. The numerical methods used in this study are detailed in \cite{viola2023high}. A summary of the numerical approach is provided in the following sections.

\subsection{Fluid Mechanics}
\label{sssec:fluidMech}
The blood velocity $\vel(\pos,t)$ and pressure $p(\pos,t)$ are governed by the incompressible Navier-Stokes equations:
\begin{equation}
    \begin{cases}\displaystyle
    \rho \left(\frac{\partial \vel}{\partial t} + \vel\cdot\nabla\vel\right) = -\nabla p + \nabla\cdot \viscstress + \mathbf{f}_\textup{IB} + \mathbf{f}_\textup{WK}\\
    \nabla\cdot\vel=0
    \end{cases}
    \label{eq:ns}
\end{equation}
where $\rho = 1000\,\,\text{Kg}/\text{m}^3$ is the constant blood density. Blood is modeled as an incompressible Newtonian fluid, so the viscous stress tensor is given by $\viscstress = \mu(\nabla \vel + \nabla \vel^T )$ with $\mu = 3.5$ \unit{\milli\pascal.s} representing the blood viscosity with a hematocrit of $40\%$ \cite{katritsis2007wss,devita2013non_newtonian}. Equations~\ref{eq:ns} are solved in dimensionless form over an Eulerian Cartesian grid using central second-order finite differences discretized on staggered grids \cite{verzicco1996finite,van2015pencil} and are marched in time using a fractional step with an explicit Adams-Bashforth method for the viscous and non-linear convective term. The non-dimensionalization process results in a Reynolds number $Re=\rho U L / \mu=3179$, where $U=0.5$ \unit{\metre\per\second} and $L=21$ \unit{\milli\metre} are the characteristic velocity and length scales used for non-dimensiolization. 
The Eulerian grid is composed of $n=n_x n_y n_z=636\times636\times726\simeq293\cdot10^{6}$ equally spaced grid points, distributed every 230 \unit{\micro\metre} in each direction, corresponding to approximately 300 million degrees of freedom for hemodynamics, about 30 million of which lie within the cardiovascular structures. This proportion is influenced by the presence of peripheral arteries and veins extending away from the heart, which reduces the percentage of active computational nodes. In more compact anatomical configurations, this percentage tends to increase. The no-slip condition on the moving wet cardiovascular tissues is imposed through the instantaneous forcing $\mathbf{f}_\textup{IB}$ using an ``Immersed Boundary Method (IBM) based on the Moving Least Square (MLS)" interpolation \cite{deTullio2016MLS,spandan_2017_potential,spandan2018fast}. As it happens in IBMs, the cardiovascular structures are immersed in the computational domain (Eulerian grid), as shown in Figure~\ref{fig:Figure_setup} and the no-slip forces $\mathbf{f}_\textup{IB}$ are applied at Lagrangian markers that are uniformly distributed over the wet cardiovascular surfaces. These surfaces are discretized using triangular elements, with their centroids serving as Lagrangian marker nodes. The hydrodynamic loads acting on the structures (to be transferred to the structural solver, detailed in the following section) are evaluated at the same Lagrangian markers located on the immersed body surface. The local force at each triangular face $\mathbf{f}^H_f$, which includes both pressure and viscous stresses, is computed as:
\begin{equation}
    \mathbf{f}^H_f = (-p \hat{\mathbf{n}}_f + \viscstress \cdot \hat{\mathbf{n}}_f)\,s\,,
    \label{eq:hydro_forces}
\end{equation}
where $s$ represents the surface area of the triangular face and $\hat{\mathbf{n}}_f$ its normal direction. These forces are then transferred to each triangle node $i$ ($\mathbf{f}^H_i$), where the structural dynamics is actually solved.\\
\indent As visible in Figure~\ref{fig:Figure_setup}, the tips of the arteries and the left ventricle representing the inlets/outlets of our cardiovascular domain do not cross the boundaries of the fluid computational domain. During the cardiac dynamics, blood is sucked from the outer volume through the mitral valve opening and propelled towards the same outer volume through the aorta. To model the impact of the truncated cardiovascular tree and obtain realistic blood pressure in the heart chambers, appropriate lumped parameters models should be used as boundary conditions for the pressure. For the descending aorta pressure we use the three element Windkessel model \cite{lantz2011windkessel} which describes the peripheral resistance and arterial compliance by solving the equivalent electrical circuit
\begin{equation}
  p^\textup{WK}_\Omega + R_dC_d\frac{dp^\textup{WK}_\Omega}{dt} = (R_d+R_p)Q^\textup{WK}_\Omega+R_dR_pC_d\frac{dQ^\textup{WK}_\Omega}{dt}\,,
  \label{eq:WK}
\end{equation}
where $p^\textup{WK}_\Omega$ is the pressure to impose as boundary condition, $Q^\textup{WK}_\Omega$ is the volumetric flowrate across the descending aorta and $R_d$, $R_p$ and $C_d$ are the resistances and capacitor parameters. These three parameters are tuned to obtain standard physiological pressure ranges inside the aorta (80-120 mmHg) \cite{westerhof2009windkessel,boccadifuoco2018impact,romarowski2018_3WK}. We use $R_d=907\times 10^{6}$ \unit{kg.m^{-4}.s^{-1}}, $R_p=11.6\times 10^{6}$ \unit{kg.m^{-4}.s^{-1}} and $C_d=7.33\times 10^{-9}$ \unit{m^4.s^2.kg^{-1}}. The pressure provided by the Windkessel model $p^\textup{WK}_\Omega$ is imposed through the volume forcing $\mathbf{f}_\textup{WK}$ in Equation~\ref{eq:ns}, which is only active in the cylindrical subdomain $\Omega$ (having outward-pointing normal vector $\hat{\mathbf{n}}_{\Omega}$), as in Figure~\ref{fig:Figure_setup}. The forcing is homogeneous inside $\Omega$ and dynamically computed at each time step using a proportional-integral-derivative controller on the error value $e_{\Omega}=(p_{\Omega}-p^\textup{WK}_{\Omega})$ (as it is done in \cite{scarpolini2025nudging}), where $p_{\Omega}$ is the average pressure measured at the inward-facing base of the cylinder $\Omega$, and $p^\textup{WK}_{\Omega}$ is the pressure obtained by solving the Windkessel model (Equation~\ref{eq:WK}). On the other hand, owing to the lower pressure loads inside the pulmonary circulation, on the pulmonary veins we set a zero pressure condition (which corresponds to sucking blood directly from the outside box).

\subsection{Structural mechanics}
\label{sssec:structMech}
\begin{figure}
  \centering
  \includegraphics{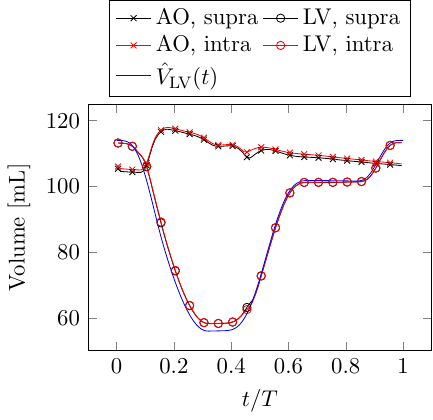}
  \caption{Volume $\hat{V}_{\text{LV}}(t)$ and surface area $\hat{A}_{\text{LV}}(t)$ laws imposed on the left ventricle using nudging algorithm \cite{scarpolini2025nudging}.}
  \label{fig:vol_law}
\end{figure}

In this work, the effect of the active ventricular contraction is prescribed using nudging to enforce a physiological volume $\hat{V}_{\text{LV}}(t)$ and surface area $\hat{A}_{\text{LV}}(t)$ law on the left ventricle \cite{scarpolini2025nudging}. The prescribed volume variation corresponds to a stroke volume of 58 mL and an ejection fraction of 51\%, values representative of a Caucasian male with a moderately small left ventricle (left ventricular end-distolic volume of 114~\unit{\ml}) \citep{petersen2016cardiacFunctionRanges}. This choice reflects the typical anatomical and functional conditions of patients undergoing aortic valve replacement where the decision between intra and supra-annular prosthesis positioning could lead, or not, to a prosthesis-patient mismatch scenario. Figure~\ref{fig:vol_law} depicts $\hat{V}_{\text{LV}}(t)$ and $\hat{A}_{\text{LV}}(t)$. The volume and area laws are imposed at an heart rate of 60 beats per minute, resulting in a period for each cycle $T=1$~\unit{\second}. These condition resembles a patient at rest.\\
\indent The dynamics of the deformable cardiovascular tissues is solved using a spring-network structural model based on the interaction potential approach \cite{fedosov2010systematic,hammer2011mass,gelder1998elastic_membranes}. A mechanical solver for thin shells is used for all biological tissues, which are discretized as triangulated surfaces. The structural model is built considering the triangulated network of springs, one for each edge. The mass of the structure is concentrated on the vertices of the triangles, uniformly distributed among them. The potential energy of the system accounts for the in-plane and bending stiffness of the tissues. In particular, the out-of-plane deformation of two adjacent triangles sharing an edge is modeled by means of a bending spring, whose elastic constant depends on the continuum elastic properties and the local tissue thickness. A detailed explanation, together with an explicit derivation of the forces acting on the mesh nodes, is given in \cite{deTullio2016MLS}.\\
\indent Several models of the elasticity of the myocardium are available in the literature, also accounting for its orthotropic properties \cite{holzapfel2001biomechanics,viola2020fsei}. Since the major focus of the study is hemodynamics, a simple linear elastic material was used in this work. The Young elastic moduli of the aortic tissue (0.4~\unit{\mega\pascal}), left ventricle (93~\unit{\kilo\pascal}) and left atrium (50~\unit{\kilo\pascal}) were chosen according to literature ranges \cite{jehl2021_LV_modulus,moireau_2009_filtering}. The wall thickness is considered uniform in the cardiac chambers and set to 2.1~\unit{\mm} for the aorta, 8.4~\unit{\mm} for the ventricle and atrium following physiological values in \cite{viola2020fsei} and clinical studies \cite{liu2015evolution,grossman1974wall,kennedy1966quantitative}. The leaflets of the biological valve have a Young modulus of 0.125~\unit{\mega\pascal}, while the mitral leaflets have 0.75~\unit{\mega\pascal}. All valve leaflets have a thickness of 1~\unit{\mm}. These values are comparable to reference literature ranges for human cardiac valves and bovine pericardium tissue \cite{stradins2004AV_mech_prop,caballero2017bovine_porcine_pericardium_mech_prop}.\\
\indent The contraction and relaxation of the left ventricle along with the passive motion of the aorta result from the dynamic balance among the inertia of the tissues $m\ddot{\mathbf{y}}$, the external hydrodynamic forces given by the fluid solver $\mathbf{f}^H$ in Equation~\ref{eq:hydro_forces}, the internal passive forces coming from the structural solver $\mathbf{f}^I$ and the nudging forces $\mathbf{f}^N$ generated to impose the volume and area profiles:
\begin{equation}
  m_i\frac{d{\mathbf{y}}_i}{dt} = \mathbf{f}_i^I + \mathbf{f}_i^H + \mathbf{f}_i^N\,,
  \label{eq:struc_dynamics}
\end{equation}
where $m_i$ is the concentrated mass of the $i$-th Lagrangian node and $\mathbf{y}_i$ is its position.\\
\indent In order to integrate Equations~\ref{eq:ns} and \ref{eq:struc_dynamics}, a loose coupling approach is used where the fluid is solved first and the generated hydrodynamic loads are used to evolve the structure, whose updated configuration is the input for the successive time step. This approach is computationally less expensive than the strong coupling counterpart but is prone to numerical instability, and a small time step (here $dt = 1$ \unit{\micro\second}) has to be used to ensure numerical stability. Additional details about this choice are discussed in \cite{viola2020fsei,verzicco_2022_EFM}.\\
\indent The numerical system resulting from the discretization of the fluid–structure interaction (FSI) problem demands substantial computational resources. To address this issue, the multiphysics solver is integrated using CUDA Fortran to harness the benefits of highly parallel multi-GPU computing. Further details on the implementation and scalability of this approach can be found in \cite{viola2022fsei_gpu}. For this study, each simulation is conducted on four Nvidia A100 40GB GPUs, with each cardiac cycle (at an heart rate of 60 bpm) advanced in approximately 38 hours of wall clock time.\\

\subsection{Hemolysis}\label{ssec:hemolysis}
The computational model enable us to measure the 3D blood velocity field with very high resolution, thus opening the way to additional hemodynamic metrics such as red blood cells (RBCs) damage. Blood damage is usually quantified by the hemolysis index $\HI$, defined as the ratio of released hemoglobin ($\Delta Hb$, in mg/100 ml) due to mechanical loading to the total hemoglobin within the RBC ($Hb$, in mg/100 ml) \cite{yu2017review_hemolysis}: $\HI = \Delta Hb/Hb$. Experimental studies under steady shear flow conditions \cite{giersiepen1990estimation} suggest that hemolysis is a function of the applied shear stress $\taueq$ and exposure time $\Delta t$:
\begin{equation}
  \HI = C \taueq^\beta \Delta t^\alpha,
  \label{eq:hemolysis_empirical}
\end{equation}
where $C$, $\alpha$ and $\beta$ are empirical parameters fitted by experiments. Given the highly unsteady nature of cardiovascular flows, subsequent studies have introduced an extension of Equation~\ref{eq:hemolysis_empirical} by integrating $\HI$ over time using an incremental damage formulation \cite{zimmer2000incremental_HI,deTullio2012mechHemolAorticValveProstheses}:
\begin{equation}
  \Delta\HI_i=\alpha C t^{\alpha-1}\taueq(t)^\beta \Delta t_i.
  \label{eq:continous_HI}
\end{equation}
In this work we use $C=1.228\cdot10^{-7}$, $\alpha=0.6606$ and $\beta=1.9918$ from \citet{zhang2011hemol_values}. This set of parameters was fitted to experiments with ovine blood exposed to shear stress between $30<\taueq<320$~\unit{\pascal} and exposure times between $0.03<t<1.5$~\unit{\second}.\\
\indent Various methods exist for estimating $\HI$, which can be broadly categorized into stress-based and strain-based approaches. In stress-based models, $\taueq$ in Equation~\ref{eq:hemolysis_empirical} is evaluated from a scalar reduction of the viscous stress tensor $\viscstress$ \cite{yu2017review_hemolysis}
\begin{equation}
  \taueq = \sqrt{\frac{1}{2}\viscstress:\viscstress}=\sqrt{\frac{1}{2}\text{tr}(\viscstress^2)}.
  \label{eq:stress_based_model}
\end{equation}
\indent However, stress-based models have two notable limitations. First, they assume that the mechanical stress experienced by RBCs is identical to that of the surrounding fluid. While this holds in steady flows, it becomes less accurate in pulsatile environments, such as the cardiovascular system, where RBCs exhibit viscoelastic behavior~\citet{Guglietta2020,guglietta2020lattice,guglietta2021loading,gurbuz2023effects,gironella2024viscoelastic,mantegazza2024red}. Second, reducing the full tensorial nature of stress to a single scalar inevitably oversimplifies the complex deformations that RBCs undergo.\\
\indent An alternative approach considers that hemolysis is not an instantaneous response to local stress but rather depends on a combination of intermediate variables, such as RBC membrane deformation and elongation. Models based on this assumption are referred to as strain-based models \cite{yu2017review_hemolysis}. A widely used strain-based model is the one by \citet{arora2004tensorial_hemolysis}, originally developed from \cite{maffettone1998ellipsoidal_drops_deformation}, in which an RBC is assumed to be ellipsoidal at all times. Its shape and orientation are described by a symmetric, positive-definite second-order morphology tensor $\deftens$, whose evolution is governed by:
\begin{equation}
  \frac{d\deftens}{dt} - \left[ \rotrate,\deftens \right] = 
    - f_1\!\left( \deftens - g(\deftens)\Id \right) 
    + f_2\!\left(\straintens, \deftens \right)
    + f_3\!\left[\vorticitytens-\rotrate, \deftens \right].
  \label{eq:arora_model}
\end{equation}
Here, $[\mathbf{A},\mathbf{B}]=\mathbf{AB}-\mathbf{BA}$ and $(\mathbf{A},\mathbf{B})=\mathbf{AB}+\mathbf{BA}$ denote the commutator and anti-commutator of second-order tensors, respectively. The fluid strain rate $\straintens$ and vorticity $\vorticitytens$ tensors -- the symmetric and anti-symmetric components of $\nabla \vel$, respectively -- are defined as $\straintens=1/2\left(\nabla \vel + \nabla \vel^T \right)$ and $\vorticitytens=1/2\left(\nabla \vel - \nabla \vel^T \right)$, and $\rotrate$ represents the rotation rate of the RBC’s frame relative to the fixed frame. This term is computed as the time derivative of the eigenvectors of $\deftens$ and $f_1$, $f_2$ and $f_3$ are model parameters calibrated to capture RBC-specific behavior: $f_1=5.0~\unit{\second^{-1}}$, $f_2=f_3=4.2298\cdot10^{-4}$ (see \cite{arora2004tensorial_hemolysis} for additional details). In this models, an effective stress acting on the RBC is defined in terms of its ``maximal" deformation $\phi=(L-B)/(L+B)$, where $L$ and $B$ are the lengths of the major and minor axes of the ellipsoid, respectively, and
\begin{equation}
  \taueq=\mu\frac{2\phi f_1}{(1-\phi^2)f_2},
  \label{eq:strain_based_stress}
\end{equation}
where $\mu$ is the fluid viscosity. $L^2$ and $B^2$ can be computed from $\deftens$ as they are its largest and smallest eigenvalues.\\
\indent Hemolysis models can also be classified based on their Eulerian or Lagrangian formulation. In Eulerian stress-based models, an Eulerian hemolysis field can be defined as $\hemolL(\pos,t)=\HI(\pos,t)^{1/\alpha}$, leading to the following transport equation \cite{yu2017review_hemolysis}:
\begin{equation}\displaystyle
  \frac{\partial \hemolL}{\partial t} + \vel\cdot\nabla \hemolL = C^{1/\alpha} \taueq^{\beta/\alpha} \left(1-\hemolL\right).
  \label{eq:hemolysis_advection}
\end{equation}
Consequently, this scalar equation, one-way coupled with the Navier-Stokes equations (Equation \ref{eq:ns}), must be discretized on the Eulerian grid and solved simultaneously. Although it is possible to derive a simular Eulerian formulation also for strain-based models \cite{dirkes2024eulerian_tensorial_hemolysis}, they are more easily implemented using Lagrangian approaches, where RBCs are tracked along their trajectories to capture cumulative deformation effects \cite{deTullio2012mechHemolAorticValveProstheses}.\\
\indent In this work, Lagrangian trajectories are thus advected from the velocity field $\vel(\pos,t)$ (assuming particles with no inertia) by integrating numerically
\begin{equation}
\begin{cases}\displaystyle
  \frac{d\mathbf{X}_i}{dt} = \vel(\mathbf{X}_i(t), t)\ , \\
  \mathbf{X}_i(t_{i0}) = \mathbf{X}_{i0},
\end{cases}
\label{eq:lagrangian_tracers}
\end{equation}
where $\mathbf{X}_i(t)$ is the position at time $t$ of the $i$-th Lagrangian particle, injected in a random position $\mathbf{X}_{i0}$ inside the cardiovascular structures, at time $t_{i0}$. The instantaneous particle's position $\mathbf{X}_i(t)$ can be used to either evaluate the fluid's stress for stress-based models (using Eq.~\ref{eq:stress_based_model}) or to evolve the RBC's deformation and compute an equivalent stress for strain-based models (using Eq.~\ref{eq:arora_model} and Eq.~\ref{eq:strain_based_stress}).



\section{Results}\label{sec:results}
\begin{figure}[h]
  \centering
  \includegraphics[width=0.49\textwidth]{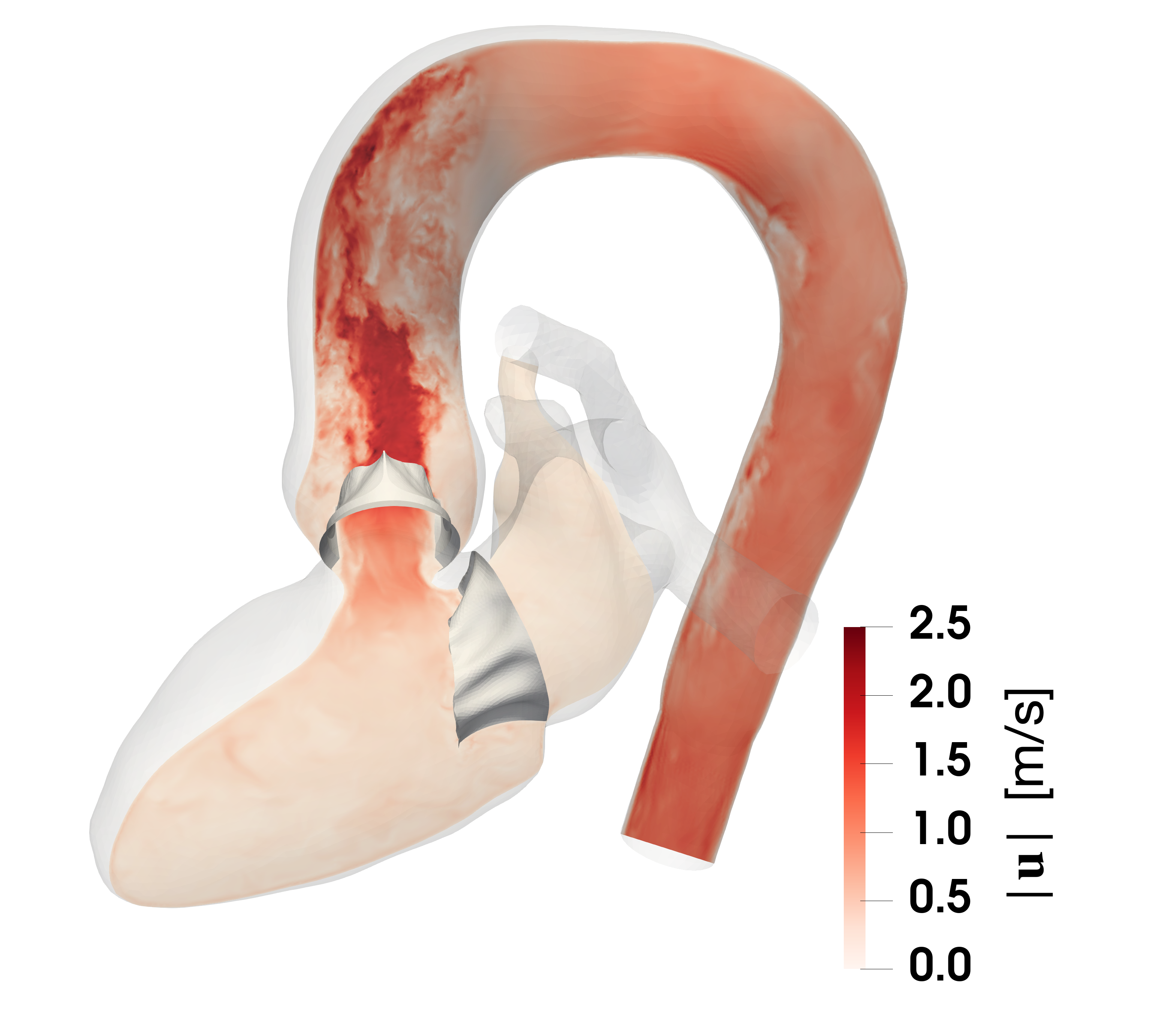}%
  \includegraphics[width=0.49\textwidth]{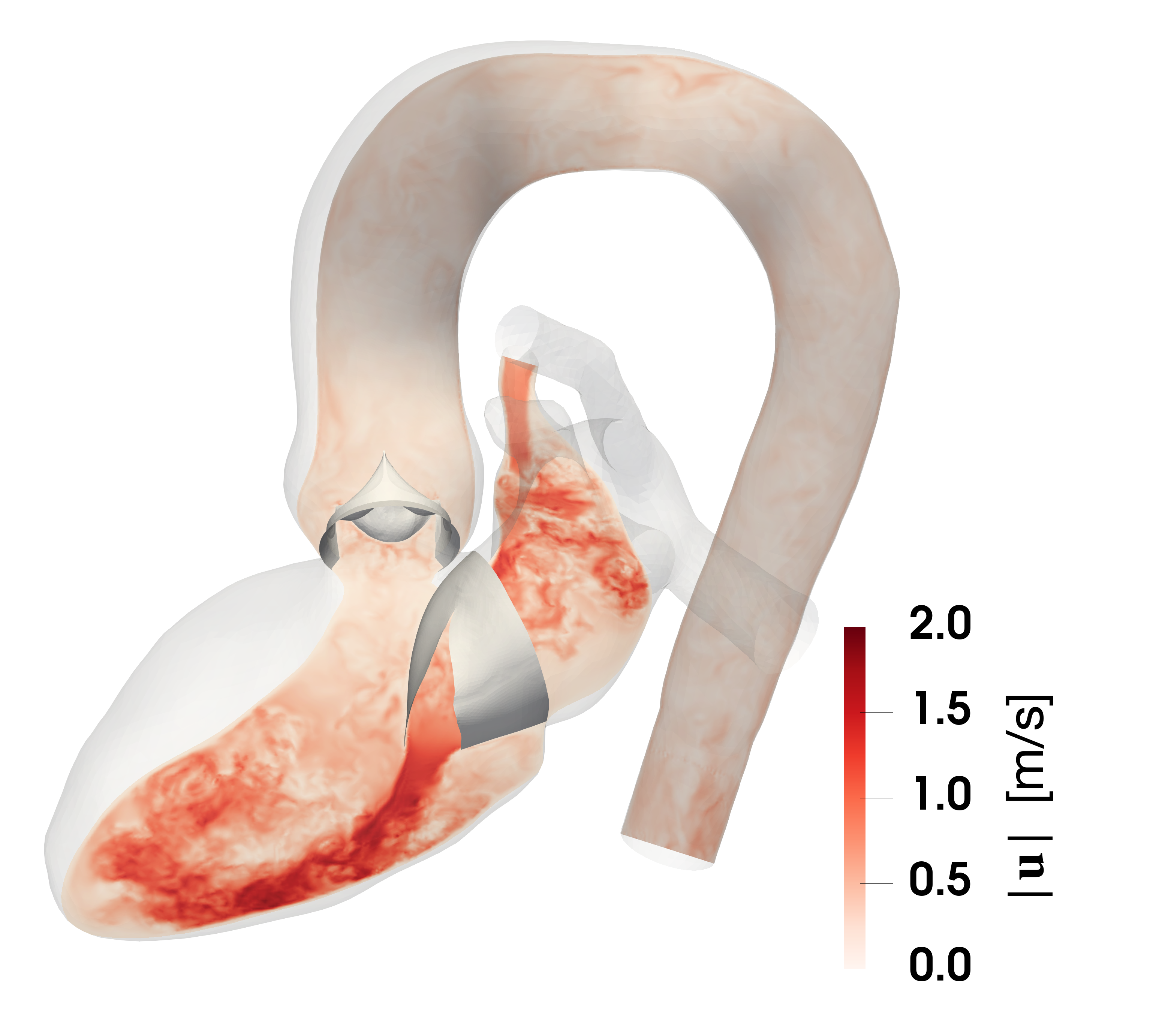}
  \caption{Instantaneous velocity magnitude distribution $|\vel(\pos,t)|$ for the supra-annular configuration at peak systole ($t/T=0.19$, left) and peak diastole ($t/T=0.58$, right). The velocity field is sampled along a cross-sectional surface positioned approximately at the midsection of each cardiac chamber, providing a detailed view of the internal hemodynamics.}
  \label{fig:global_hemol}
\end{figure}

The results for the two valve configurations are presented and compared in this section. Each simulation was run for six cardiac cycles. Since the initial conditions assume zero pressure and velocity fields ($p(\pos, t=0)=0$ and $\vel(\pos, t=0)=0$), the first two cycles are discarded to eliminate transient effects and ensure a physiologically realistic regime state. To provide an overview of the computational setup, Figure~\ref{fig:global_hemol} displays the instantaneous velocity magnitude distribution, $|\vel(\pos,t)|$, within the cardiac structures for the supra-annular configuration. The velocity field is sampled along a cross-sectional surface positioned approximately at the midsection of each cardiac chamber, providing a detailed view of the internal flow structures at two characteristic time instants of the cardiac cycle: peak systole ($t/T=0.19$, left) and peak diastole ($t/T=0.58$, right). During systole, the contraction of the left ventricle generates a strong aortic jet as blood is ejected through the prosthetic valve into the ascending aorta. The figure captures a two-dimensional slice of this high-velocity jet and its impingement on the ascending aortic wall. As the flow progresses along the aortic arch, it develops into a more structured pattern with minimal disturbances, maintaining relatively stable transport into the descending aorta. Following the systolic peak, ventricular contraction decelerates, reducing the velocity of the aortic jet until the pressure gradient between the left ventricle and aorta reaches zero. At this point, after a short isovolumic time interval, ventricular relaxation begins, marking the transition to diastole. The left ventricle expands, generating a counter pressure gradient drawing blood from both the left atrium and, momentarily, from the aorta. This flow reversal induces the rapid closure of the aortic valve while the mitral valve opens, initiating ventricular filling. The right panel of Figure~\ref{fig:global_hemol} captures the peak diastolic inflow, characterized by the mitral jet, which -- though less intense than the aortic jet due to the larger orifice area -- produces significant flow acceleration. The jet impinges on the ventricular wall, which, due to its anatomical features, generates a large scale vortex that persists for large parts of the cardiac cycle. Meanwhile, oxygenated blood from the pulmonary veins enters the left atrium, preparing for the next cycle. These key hemodynamic features confirm that the computational setup effectively replicates physiological cardiac dynamics, ensuring a meaningful comparison between the two valve configurations.

\subsection{Wiggers diagram}\label{sec:wig}
\begin{figure}[h]
  \centering
  \includegraphics{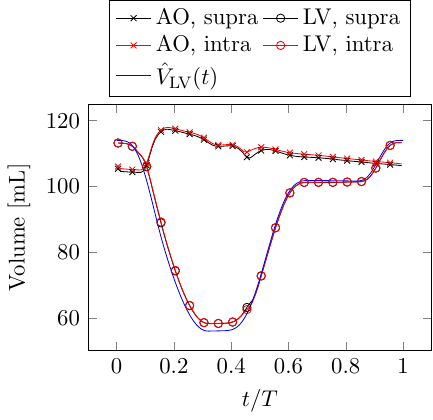}%
  \includegraphics{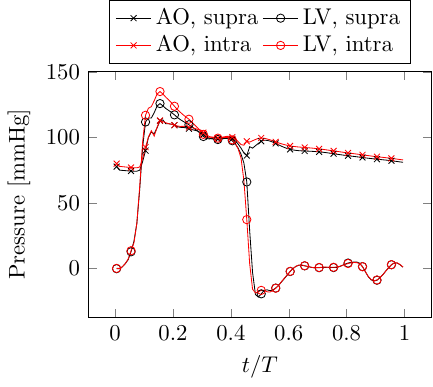}
  \caption{Wiggers diagram. Aortic (AO) and left ventricular (LV) volumes are shown on the left panel, together with the imposed volume law $\hat{V}_{\text{LV}}(t)$ in blue. Pressures of the corresponding cardiac chambers are shown on the right.}
  \label{fig:wiggers}
\end{figure}

Figure~\ref{fig:wiggers} presents the Wiggers diagram, showing volumes and pressure from the computational model for both supra-annular and intra-annular valve configurations. The first panel illustrates that, by employing nudging, the desired volumetric flow rate $\hat{V}_{\text{LV}}(t)$ was successfully prescribed throughout the cardiac cycle, ensuring an identical stroke volume for both valve positions. This guarantees that any hemodynamic differences observed between the two configurations are solely attributed to valve positioning rather than variations in ventricular ejection. The aortic volume curve reflects the characteristic increase in aortic volume during systole, driven by the rise in intraluminal pressure and the subsequent deformation of the aortic wall due to fluid-structure interaction. Notably, this volumetric response does not exhibit significant differences between the intra-annular and supra-annular cases, suggesting that global aortic compliance remains comparable across configurations.\\
\indent The second panel of Figure~\ref{fig:wiggers} shows the ventricular and aortic pressures. The implementation of Windkessel boundary conditions effectively maintains aortic pressure within the physiological range of approximately 80–120~\unit{\mmHg}, accurately capturing the resistive and capacitive characteristics of the peripheral circulation. The aortic pressure curve shows the characteristic pattern throughout the cardiac cycle, reflecting the phases of ventricular systole and diastole. During systole, as the left ventricle contracts and the aortic valve opens, blood is rapidly ejected into the aorta, causing a steep pressure increase. As ventricular ejection slows, pressure begins a gradual drop lasting till the end of diastole, when the diastolic pressure value of $\sim80$~\unit{\mmHg} is attained. At the end of systole, the aortic valve closes, generating a distinct bump (dicrotic notch) on the pressure curve. This jump marks the brief pressure rise due to elastic recoil of the aortic walls and retrograde flow momentarily pushing against the closed valve (false regurgitation). These features are present in both valve configurations and do not evidence significant differences. Aortic distensibility $\AD$, a standard parameter measured in clinical studies, quantifies the percentage change in aortic diameter (during a cardiac cycle) per unit of pressure variation:
\begin{equation}
  \AD = 2\,\,\frac{\left(D_\text{sys}-D_\text{dia}\right)/D_\text{dia}}{p_\text{sys}-p_\text{dia}},
\end{equation}
where $D_\text{sys}$ and $D_\text{dia}$ are the aortic diameters in systole and diastole, respectively, and $p_\text{sys}$ and $p_\text{dia}$ are the corresponding pressures. These measurements are performed on the ascending aorta, approximately 4~\unit{\cm} after the aortic valve. In our simulations, $\AD$ is approximately 4.3~\unit{\cm\squared\per\dyne} for both configurations, in agreement with physiological values reported in \cite{stefanadis1990distensibility}. $\AD$ does not vary significantly due to its dependence on aortic pressures, which are very similar between the two configurations. Contrarily, the primary distinction caused by deployment position is observed in ventricular pressure. The intra-annular configuration requires a higher left ventricular pressure -- approximately 10~\unit{\mmHg} more than the supra-annular one, to ensure the same flowrate across the valve. This difference highlights the hemodynamic advantage of supra-annular positioning in reducing the ventricular workload. 

\subsection{Valve orifice area and Transvalvular pressure drop}\label{sec:goa_tpd}
\begin{figure}[h]
  \centering
  \includegraphics{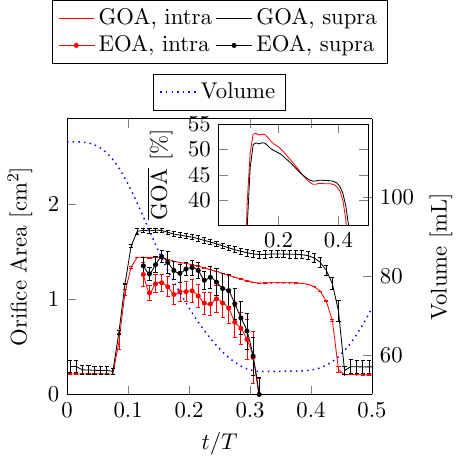}%
  \includegraphics{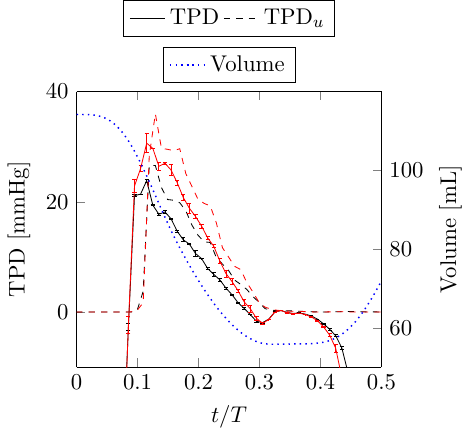}%
  \caption{Valve orifice area ($\GOA$ and $\EOA$) and trans-valvular pressure drop ($\TPD$ and $\TPD_u$), both reported in terms of direct and indirect measurements (the latter corresponding to clinical practice). The inset plot of the first panel shows the normalized geometric orifice area $\overline{\GOA}$. The curves are phase-averaged over four cardiac cycles. The underlyng dotted curve in both panels indicate the ventricular contraction and expansion during the heatbeat.}
  \label{fig:GOA_TPD}
\end{figure}

The first metric related to valve performance is the valve orifice area, specifically the Geometric Orifice Area ($\GOA$). The latter is also referred to as anatomical aortic valve area. $\GOA$ represents the minimal cross-sectional area through which the flow is forced to pass, providing a direct geometric measure of the valve opening. To evaluate this area relative to the prosthesis size, an additional normalized metric is introduced as $\overline{\GOA}=\GOA/(\pi(D_I/2)^2)$, where $D_I$ is the maximum internal diameter of the prosthesis (Figure~\ref{fig:Figure_setup}). In clinical practice, $\GOA$ is typically measured either through transthoracic echocardiography (TTE) or transoesophageal echocardiography (TEE). However, imaging artifacts such as shadows or reverberations, particularly in calcified stenotic valves, limit the identificability of the orifice, especially with TTE. Ensuring that the minimal orifice area is identified, rather than the larger area proximal to the leaflet tips, can become difficult, leading to high uncertainties \cite{baumgartner2017ESC_AVS_assessment_echocardiography}. Computational models, in contrast, provide a precise geometric representation, enabling a more accurate assessment of valve function.\\
\indent Due to the limitations of imaging techniques, clinicians often rely on the ``Effective Orifice Area" ($\EOA$), an indirect measurement of the orifice area, estimated using the continuity equation:
\begin{equation}
  \EOA = \frac{Q_\text{AV}}{u_{\text{jet}}},
\end{equation}
where $Q_\text{AV}$ is the volumetric flowrate across the aortic valve, and $u_{\text{jet}}$ is the velocity of the aortic jet during ventricular systole. Since $\EOA$ is derived from the central jet velocity, it primarily reflects the region of highest flow velocities, leading to a slightly smaller value compared to the $\GOA$, which accounts for the full anatomical opening. Both $Q_\text{AV}$ and $u_{\text{jet}}$ vary over time, leading to a time-varying $\EOA$ throughout ventricular systole. However, $\EOA$ becomes ill-defined when these quantities approach zero, leading to an artificial drop to zero before the valve closes.\\
\indent Another critical performance metric is the transvalvular pressure drop $\TPD$, representing the pressure difference across the prosthetic valve during the systolic phase. Medical literature usually name this metric ``transvalvular pressure gradient", even tough there is no differentiation with respect to space. This metric is crucial as it directly reflects the resistance imposed by the valve on blood flow. The latter is closely linked to the $\GOA$ -- intuitively, a smaller $\GOA$ results in higher flow resistance, leading to an increased pressure drop and greater workload on the heart. $\TPD$ is defined as the difference between the pressure upstream $p_U$ and downstream $p_D$ of the valve:
\begin{equation}
  \TPD=p_U-p_D.
  \label{eq:tpd}
\end{equation}
In our numerical setup, $p_U$ and $p_D$ are extracted from the Eulerian pressure field $p(\pos,t)$ within two measurement regions located 2~\unit{\cm} upstream and downstream the aortic valve. These pressures are averaged over cubic regions with an edge length of 4~\unit{\mm} to reduce local fluctuations. In clinical settings, direct pressure measurements are uncommon due to their invasive nature. Instead, $\TPD$ is estimated indirectly using the velocity of the aortic jet downstream of the valve:
\begin{equation}
  \TPD_{u} = 4u_{\text{jet}}^2,
  \label{eq:TPD_indirect}
\end{equation}
where $u_{\text{jet}}$ must be measured in \unit{\meter\per\second} and $\TPD_u$ turns out to be in \unit{\mmHg} (with the `magic' factor 4 accounting for all dimensions and unit conversions) \cite{baumgartner2017ESC_AVS_assessment_echocardiography}. This indirect estimation of $\TPD$ is widely used in clinical studies and is believed to be based on the Bernoulli equation. These issues further highlight the advantages of our computational approach when evaluating valve efficiency metrics. The velocity $u_{\text{jet}}$ is measured following the same procedure as $p_D$ (location and spatial averaging), for both $\EOA$ and $\TPD_u$.\\
\indent Figure~\ref{fig:GOA_TPD} shows $\GOA$, $\EOA$, $\overline{\GOA}$, $\TPD$ and $\TPD_u$ during the ventricular systole. Each quantity is phase-averaged along four cardiac cycles and the errorbars correpond to the standard deviation. At peak systole, the intra-annular valve reaches a $\GOA$ of approximately $1.45$~\unit{\cm\squared}, whereas the supra-annular valve, benefitting from its larger initial size, attains a $\GOA$ of approximately $1.7$~\unit{\cm\squared}. This corresponds to a $\sim17\%$ increase, which has noticeable implications for hemodynamics, as discussed in the next sections. The normalized orifice area $\overline{\GOA}$ (inset of Figure~\ref{fig:GOA_TPD}) reveals a slightly higher efficiency for the intra-annular valve (~53\% vs ~51\% for the supra-annular configuration), though clinical outcomes primarily depend on absolute $\GOA$ rather than relative efficiency. It is noteworthy that the peak values for both configurations occur at the same time instant. In both cases, the valve requires approximately $120$~\unit{\ms} to reach its maximum opening state. This phase is followed by a gradual decrease in $\GOA$ during the systolic deceleration phase, eventually stabilizing for approximately $100$~\unit{\ms} during the diastolic plateau, a phase in which ventricular motion is minimal. Subsequently, the valve undergoes rapid closure, completing this transition in approximately $80$~\unit{\ms}.\\
\indent From Figure~\ref{fig:GOA_TPD} it is also evident that the supra-annular configuration shows larger cycle-to-cycle variations in $\GOA$ (highlighted by the error bars). This variability primarily results from increased cycle-to-cycle differences in the opening behavior of the right and left coronary cusps (RCC and LCC) leaflet. Specifically, we observe that the free edge of the valve, a region often affected by flow instabilities, exhibits small, non-monotonic variations across beats in the supra-annular case. These subtle variations in leaflet opening contribute to the cycle-to-cycle differences in the computed $\GOA$. This phenomenon might be due to the interaction between leaflet dynamics and the surrounding unsteady flow field. In the supra-annular configuration, indeed, the LCC and RCC leaflets are positioned farther from the sinus wall, making them more susceptible to the central jet and associated turbulence features. In contrast, the intra-annular configuration positions the leaflets closer to the sinus wall, where boundary-layer effects and local flow confinement likely stabilize the leaflet motion. A deeper analysis would be beneficial to fully characterize this behavior.\\
\indent On the other hand, the $\EOA$ shows peak values of approximately $1.17\pm0.09$~\unit{\cm\squared} and $1.45\pm0.06$~\unit{\cm\squared} for the intra and supra configurations, respectively. These values are smaller than their geometric counterparts, as anticipated previously, as they reflect the region of highest flow velocities. However, they confirm the same trend observed with $\GOA$, favoring the supra-annular position. Additionally, as observed previously, $\EOA$ drops artificially to zero before the valve closes, and later becomes ill-defined. Consequently, its values are shown only until $Q_\text{AV}$ remains positive.\\
\indent Importantly, $\EOA$ values are used to assess PPM through the ``indexed $\EOA$'', which is obtained by normalizing $\EOA$ to the patient's body surface area ($\BSA$). Given the patient's small aortic annulus, a $\BSA$ value of 1.84~\unit{\meter\squared}--corresponding to one standard deviation below the adult male mean \cite{tikuisis2001BSA_from_CT}--is used. Based on this, the indexed $\EOA$ is 0.63 and 0.79~\unit{\cm\squared}/\unit{\meter\squared} for the intra-annular and supra-annular configurations, respectively. Indexed $\EOA$ values below 0.85 (0.65)~\unit{\cm\squared}/\unit{\meter\squared} define the thresholds for moderate (severe) PPM \cite{pibarot2006PPM}. These results indicate that the intra-annular configuration would result in moderate-to-severe PPM, whereas the supra-annular configuration would place the patient almost outside the PPM region.\\
\indent Figure~\ref{fig:GOA_TPD} illustrates the temporal evolution of $\TPD$ and $\TPD_u$ during systole for both deployment configurations. At systolic peak, the intra-annular valve exhibits a maximum $\TPD$ of approximately $\sim30$~\unit{\mmHg}, whereas the supra-annular valve reaches a lower maximum value of around $\sim23$~\unit{\mmHg}. This  reduction in peak pressure drop for the supra-annular configuration can be attributed to its larger orifice area, which facilitates improved blood flow dynamics and reduces resistance across the valve. In clinical studies, $\TPD$ is often reported in terms of the ``mean transvalvular pressure drop", which represents the average of $\TPD$ over systole. For the intra-annular and supra-annular configurations, this value is 17.7~\unit{\mmHg} (18.2 for the indirectly measured one) and 12~\unit{\mmHg} (12.3 for the indirectly measured one), respectively, further highlighting the superior hemodynamic efficiency of the supra-annular design. These values agree with tyipical values measured in clinical practice, as reported in \citet{baumgartner2017ESC_AVS_assessment_echocardiography} for both healthy and mildly stenotic aortic valves. Interestingly, despite its simplified formulation (Equation~\ref{eq:TPD_indirect}), the velocity-based $\TPD_u$ closely approximates the direct numerical measurement, supporting its practical utility in clinical practice.

\subsection{Wall shear stress}\label{sec:wss}
\begin{figure*}[!ht]
  \centering
  \includegraphics[width=1\textwidth]{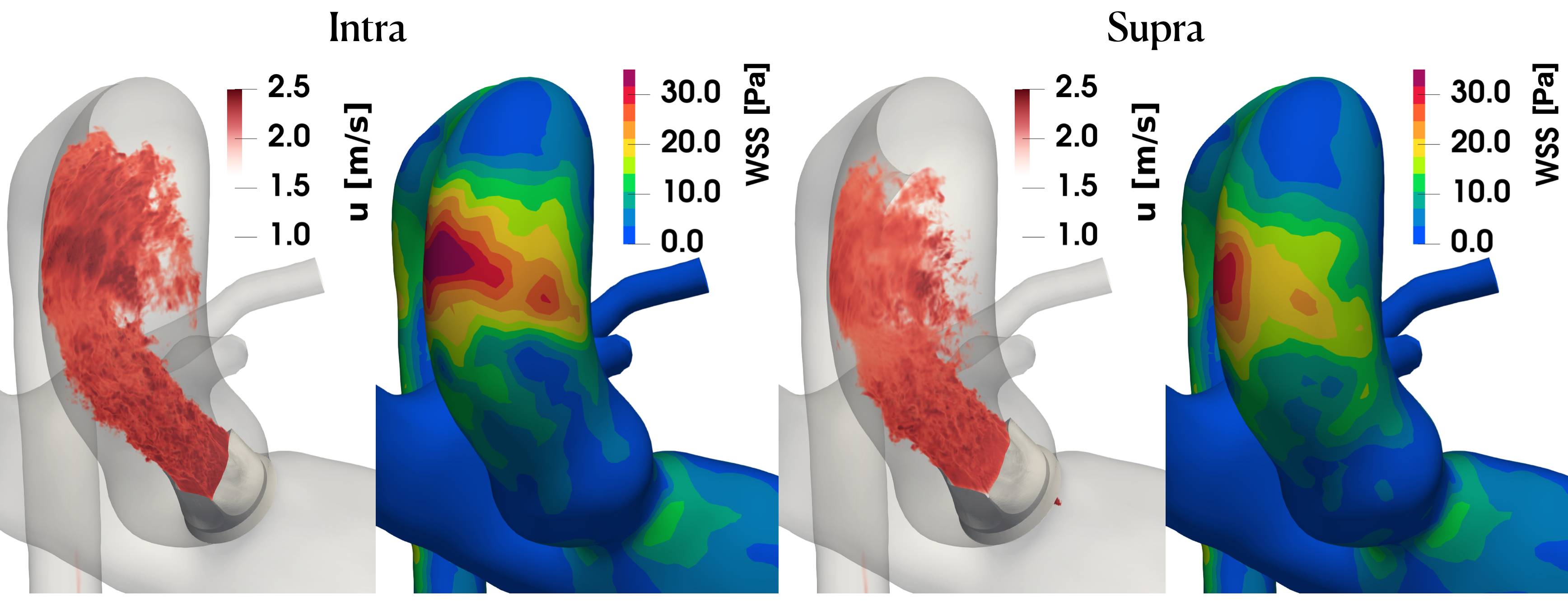}
  \caption{Aortic jet impingement on the ascending aorta wall. Blood velocity magnitude $|\vel|$ and $\WSS$ during systolic peak, compared between intra and supra configurations. The curves are phase-averaged along four cardiac cycles.}
  \label{fig:Figure_wss_vel}
\end{figure*}

\begin{figure}
\centering
\includegraphics{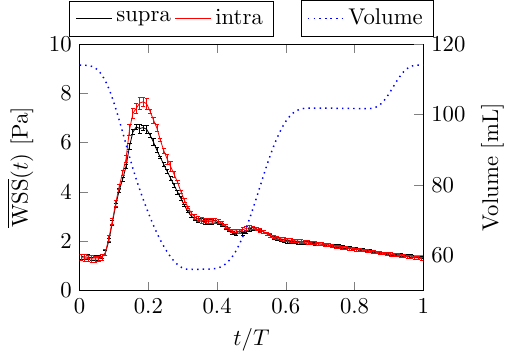}%
\caption{Wall shear stress averaged across the ascending aorta, $\aWSS(t)$, in the intra (red) and supra (black) case. The curves are phase-averaged along four cardiac cycles and errorbars are computed as the cycle-to-cycle standard deviation.}
\label{fig:wss_asc}
\end{figure}

Wall Shear Stress (\WSS) is another key hemodynamic quantity to assess tissue damage defined from the viscous stress tensor:
\begin{equation}
  \mathbf{T}=\viscstress\cdot\hat{\mathbf{n}}; \quad\quad \WSS=|\mathbf{T}-\left(\mathbf{T}\cdot\hat{\mathbf{n}}\right)\hat{\mathbf{n}}|,
  \label{eq:wss}
\end{equation}
where $\hat{\mathbf{n}}$ is the wall normal direction. $\WSS$ represents the tangential component of the force exerted by the blood flow on the cardiovascular structures. Given its dependence on velocity gradients near the wall, $\WSS$ plays a crucial role in vascular biology, influencing endothelial cell function and mechanotransduction processes. Abnormal WSS distributions have been associated with vascular remodeling and pathological processes such as atherosclerosis and aneurysm formation \cite{verzicco_2022_EFM}. In the aorta, $\WSS$ is closely linked to the impingement of the aortic jet on the ascending wall, which is largely dictated by the flow characteristics induced by the biological or prosthetic valve. Figure~\ref{fig:Figure_wss_vel} presents the velocity magnitude $|\vel|$ at the systolic peak alongside the corresponding $\WSS$ distribution. The intra-annular configuration generates a more concentrated aortic jet with higher velocities, leading to stronger velocity gradients near the aortic wall. This results in an extended region of elevated WSS values ($\sim30-40$ \unit{\pascal}) compared to the supra-annular configuration ($\sim15-25$ \unit{\pascal}), where the jet impingement area is less focused. Such $\WSS$ values are comparable to those found in the shear layer of the aortic jet (around $10-40$~\unit{\pascal}, depending on the valve deployment location) and are therefore highly relevant for assessing hemolysis risk, as detailed in the next section. This finding highlights the limitations of simplified aortic models, such as straight-tube approximations commonly used in valve performance studies \cite{deTullio2012mechHemolAorticValveProstheses,corso2024valve_architecture,gallo2022valve_performance_and_helicity,becsek2020turbulent_valve_spectra,ferrari2024comparison}. Such models fail to capture the physiological complexity of cardiovascular flows, particularly the high-shear regions near the jet impingement area. A more physiologically accurate representation, such as the curved-tube models employed in \cite{azadani2017blood_stasis,zingaro2024valve_FSI_4dFlow}, may offer a more appropriate compromise between realism and computational feasibility.\\
In order to better compare the intra- and supra-aortic deployment of the prosthetic valve, we compute the spatially averaged $\WSS$ over the ascending aorta:
\begin{equation}
  \aWSS(t)=\frac{1}{A_{\text{aAO}}}\int_{\text{aAO}} \!\!\WSS(\pos,t) \,ds,
  \label{eq:average_wss}
\end{equation}
where $\aAO$ stands for ascending aorta and $A_{\aAO}$ is its surface area. As shown in Figure~\ref{fig:wss_asc}, the intra-annular configuration consistently exhibits higher $\aWSS(t)$ during the systolic peak, with an increase of approximately 1~\unit{\pascal}. This difference highlights the significant impact of valve deployment on near-wall hemodynamics, which may have important implications for long-term vascular adaptation and remodeling.

\subsection{Hemolysis}\label{sec:hemol}
To evaluate hemolysis associated with the two different valve deployment locations, we employ both Lagrangian and Eulerian formulations. Results are presented and compared between different hemolysis models in terms of the Lagrangian approach, as it enables the easy implementation of all models, including the strain-based realistic RBC deformation models, as discussed in section~\ref{ssec:hemolysis}. The Eulerian formulation, on the other hand, provides important insights into the spatial distribution of hemolysis, allowing for the assessment of high-shear regions where RBC are most likely to be deformed and ruptured.\\
\indent Lagrangian tracers are injected into the left ventricle, and their trajectories are tracked by solving Equation~\ref{eq:lagrangian_tracers} until they reach the descending aorta. The initial position of the particles, $\mathbf{X}_{i0}$, are randomly assigned within a cubic region of edge length L=2~\unit{\cm} centered inside the ventricle. To maintain a constant number of N=10,000 particles within the cardiovascular domain, each time a particle exits the aorta, a new one is injected at a randomly selected position within the defined cube. Stress, $\taueq^i(t)$, and hemolysis, $\HI^i(t)$, are evaluated for each RBC (i.e., the $i$-th tracer), based on both stress- and strain-based models using Equation~\ref{eq:stress_based_model}~or~\ref{eq:strain_based_stress}. The latter are integrated starting from inside the left ventricle, 0.5~\unit{\cm} upstream of the aortic valve, and evolved in time until they exit the descending aorta. Time evolution of $\HI$ is integrated using Equation~\ref{eq:continous_HI}.\\
\begin{figure*}[!ht]
  \centering
  \begin{subfigure}[b]{0.35\textwidth}
    \label{fig:Label-figura1}
    \includegraphics[width=\textwidth]{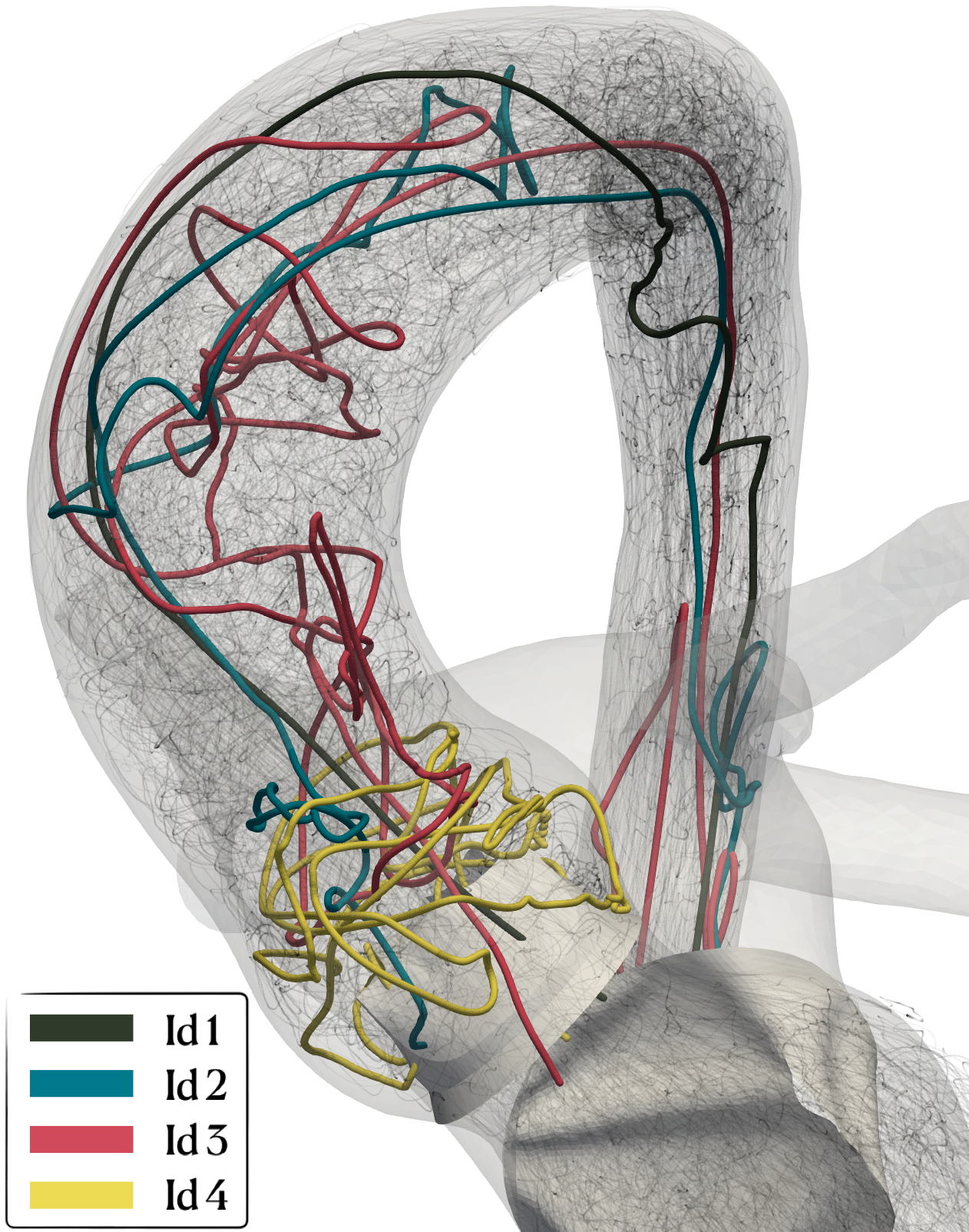}
  \end{subfigure}
  \begin{subfigure}[b]{0.55\textwidth}
    \label{fig:Label-figura2}
    \includegraphics{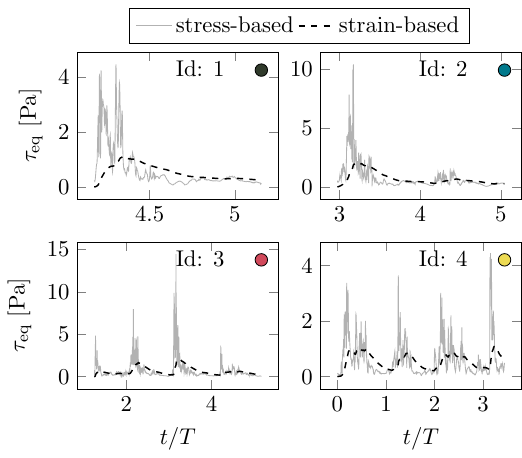}%
  \end{subfigure}
  \caption{Lagrangian trajectories $\mathbf{X}_i(t)$ inside the aorta with the bio-prothesic valve mounted in the supra location. 400 trajectories are shown in transparent black. Four tracers are highlighted with ticker colored lines, and their equivalent stress $\taueq^i(t)$ is shown on the right panel. $\taueq^i(t)$ is reported for both stress-based (solid line) and strain-based (dashed line) models.}
  \label{fig:lagr_trajs}
\end{figure*}
\indent Figure~\ref{fig:lagr_trajs} illustrates four representative Lagrangian trajectories within the aorta in the supra-deployment configuration. The left panel depicts the cardiovascular structures in transparent gray and the trajectories $\mathbf{X}_i(t)$ of the four particles in colors. Additionally, 400 tracers (out of the total 10,000) are overlaid in transparent black to provide insight into the spatial distribution of the sampled particles. The right panel shows the equivalent stress $\taueq^i(t)$ for these four tracers as predicted by both stress-based and strain-based models. Clearly, different Lagrangian particles may have significantly different paths within the aorta.\\
\indent The first trajectory (id=4, black) follows the main flow direction, impinging on the ascending aortic wall before continuing along the aortic arch in a swirling motion and exiting the vessel within approximately one cardiac cycle. The second trajectory (id=2, green) follows a more intricate path, exhibiting small-scale unsteady deviations and swirling regions. These features lead to an extended residence time of approximately two cardiac cycles. After the initial systolic phase, the trajectory briefly reverses direction, displaying a minor retrograde motion toward the valve, before ultimately exiting the vessel during the second systolic phase. The third trajectory (id=3, magenta) represents one of the most complex cases observed, with a prolonged residence time spanning four cardiac cycles. Entering the aorta during late systole, the particle undergoes a weak acceleration, insufficient to propel it out of the ascending aorta. As a result, it becomes entrapped within decaying diastolic vortices for three consecutive cycles. During this extended residence time, the particle remains in highly unsteady flow regions and is subjected to substantial velocity gradients, with $\taueq^i(t)$ reaching values in the range of 10--15 \unit{\pascal}. In the descending aorta, the particle eventually exits after one additional cycle, following recirculating flow structures in this region as well. The final trajectory analyzed (id=4, yellow) exhibits a particularly distinct behavior: it remains confined within the recirculating region of the Valsalva sinuses for over three cardiac cycles. During this period, the particle remains trapped in relatively stable vortices, experiencing lower velocity gradients that persist for an extended duration.\\
\indent Interestingly, the four particles follow very different paths and remain inside the aorta for very different time periods. Stress peaks, however, always coincide with velocity bursts during the systolic ejection, which lead to different values of $\taueq^i$, depending on the precise particle path and hemolysis model. The stress-based model predicts fluid stress fluctuations with an intermittent nature, characterized by high peak values sustained over short time intervals, very different from path to path. In contrast, the strain-based model exhibits a delayed response due to the nonzero loading time, which postpones RBCs deformation relative to $\viscstress$, and creates more similar $\taueq^i(t)$ values between different trajectories. Additionally, the deformation persists for a longer duration due to the relaxation time parameter $f_1=200$ \unit{\ms}, as defined in Equation~\ref{eq:arora_model}.\\
\begin{figure}
  \centering
  \includegraphics{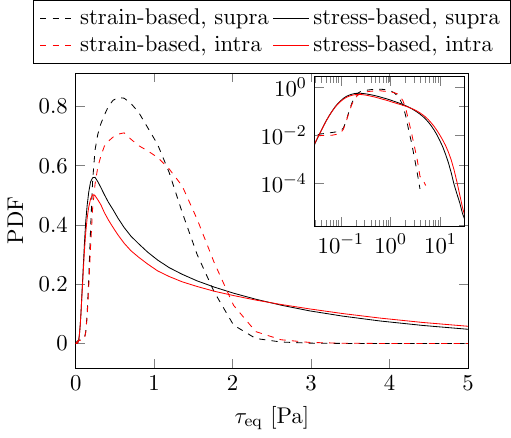}%
  \caption{PDF of $\taueq$ for both the stress- and strain-based hemolysis models computed across 10,000 Lagrangian particles inside the aorta, spanning four cardiac cycles. Black and red curves denote the supra and intra deployment locations, respectively. The inset plot shows the same PDFs in logarithmic scale. Both stress- and strain-based models are evaluated in the Lagrangian approach to be compared.}
  \label{fig:pdf_stress}
\end{figure}
\indent To accurately capture the differences between the two prosthetic deployment locations, the analysis of only a few Lagrangian trajectories is insufficient. A more statistically robust approach involves examining the probability distribution function (PDF) of $\taueq$, computed over a large ensemble of particles. Figure~\ref{fig:pdf_stress} shows the PDF of $\taueq(t)$ along the path of 10,000 tracers, injected over four cardiac cycles. For both stress-based and strain-based models, lower values of $\taueq$ are more prevalent in the supra-deployment configuration. Conversely, the intra-deployment configuration exhibits longer distribution tails, indicating a higher probability of elevated hemolysis values. An additional observation is that the stress-based PDFs are sharply peaked around $\taueq\simeq0.2$~\unit{\pascal}, but feature extensive tails that capture spikes in fluid viscous stress $\viscstress$. High values, such as $\taueq\simeq15$~\unit{\pascal}, suggest the occurrence of significant stress bursts, as already highlighted in Figure~\ref{fig:lagr_trajs}, around the systolic aortic jet region. In contrast, the strain-based model exhibits a probability peak at approximately $\taueq\simeq0.7$~\unit{\pascal}. However, its distribution tails decline much more rapidly, effectively capping around 4\unit{\pascal}. This suggests that the strain-based approach attenuates extreme stress events due to the time needed to deform the particle.\\
\indent Another statistical approach is ensemble averaging, which can be employed to estimate the mean stress and $\HI$ within the aorta at each time instant. We define ensemble averages, denoted by $\langle \cdot \rangle$, as statistical means computed over the entire set of Lagrangian particles. To implement this approach, we discretize time into 100 small intervals per cardiac cycle and compute the average values of $\taueq^i$ and $\HI$ within each interval. Specifically, for $\taueq$, the ensemble average is defined as:
\begin{equation}
  \langle\taueq\rangle(t_j) = \sum_{i, t\in[t_j-\Delta t/2, t_j+\Delta t/2)} \taueq^i(t)
  \label{eq:ensemble_average}
\end{equation}
where $\Delta t=T/100$ is the half-length of each time interval.\\
\begin{figure}
\centering
\includegraphics{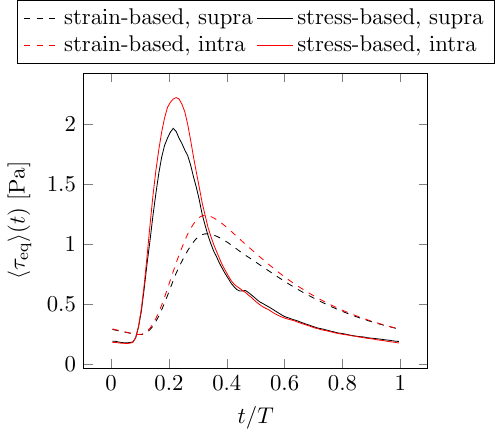}%
\includegraphics{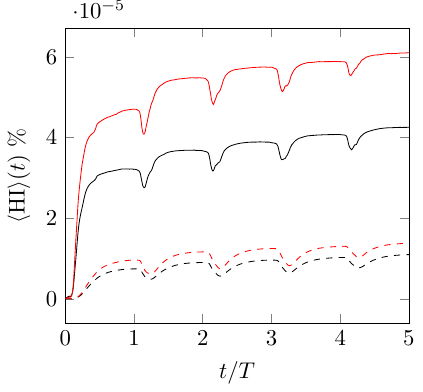}
\caption{Ensemble averaged shear stress $\langle\taueq\rangle(t)$ and hemolysis index $\langle\HI\rangle(t)$ across 10,000 Lagrangian particles in the aorta. The quantities are shown computed with both stress- and strain-based hemolysis models implemented in a lagrangian approach. $\langle\taueq\rangle(t)$ is phase-averaged along four cardiac cycles, while for $\langle\HI\rangle(t)$ the entire evolution is shown.}
\label{fig:ensemble_average_stress_HI}
\end{figure}
\indent Figure~\ref{fig:ensemble_average_stress_HI} depicts the evolution of both $\langle \taueq \rangle(t)$ and $\langle \HI \rangle(t)$. 
For the stress-based model, $\langle \taueq \rangle(t)$  peaks during systole, reaching approximately ten times its diastolic value. In contrast, the strain-based model shows a delayed peak (100~\unit{\ms}) due to the non-instantaneous nature of RBC deformation. This deformation inertia reduces peak stress to about four times its minimum value, compared to the stress-based model. Additionally, during diastole, the strain-based $\taueq(t)$ remains slightly elevated (0.3 vs 0.2~\unit{\pascal}) due to the relaxation time of RBC deformation.\\
\indent Comparing the intra- and supra-annular valve positions, the average value of $\taueq$ is higher for the intra configuration, as already suggested by the PDFs. The difference between the two is around 0.2~\unit{\pascal} for the strain-based model and 0.4~\unit{\pascal} for the stress-based one. In terms of $\langle\HI\rangle$, the two configurations differ of $\sim0.4\cdot10^{-5}\%$ and $\sim2\cdot10^{-5}\%$ for the two models. It is interesting to notice that even if Equation~\ref{eq:continous_HI} predicts a only-increasing $\HI$, the fact that fresh blood ($\HI=0$) enters the aorta at each period, makes $\langle\HI\rangle(t)$ slightly decrease during the initial phase of each systole.\\
\indent Additional insights into the hemolysis mechanism are illustrated in Figure~\ref{fig:Figure_HI_systole}, which shows the Eulerian stress-based hemolysis distribution at the peak of systole. The results clearly indicate that RBC rupture is mostly pronounced within the shear layer surrounding the aortic jet and on the jet impingement region, where viscous shear stresses $\viscstress$ reach their highest values. This mechanism helps explain the longer tails in the probability density function (PDF) of viscous stress for the intra-annular configuration, as shown in Figure~\ref{fig:pdf_stress}. Once generated, hemolysis is transported by the flow through advection and diffusion, leading to a gradual smoothing of the concentration field until a nearly homogeneous distribution is observed after valve closure. Additionally, it is interesting to notice the ``fresh blood" inflow coming from the ventricle, which replaces part of the damaged blood during each cardiac cycle inside the aorta. Also the Eulerian hemolysis model outlines an higher $\HI$ for the intra-annular configuration, both in the shear layer and on the accumulated hemolysis inside the aortic root, ascending aorta and aortic arch. All these insights are clearly shown by the Eulerian formulation, highlighting its strengths with respect to the Lagrangian one.
\begin{figure}[!h]
  \centering
  \includegraphics[width=0.7\textwidth]{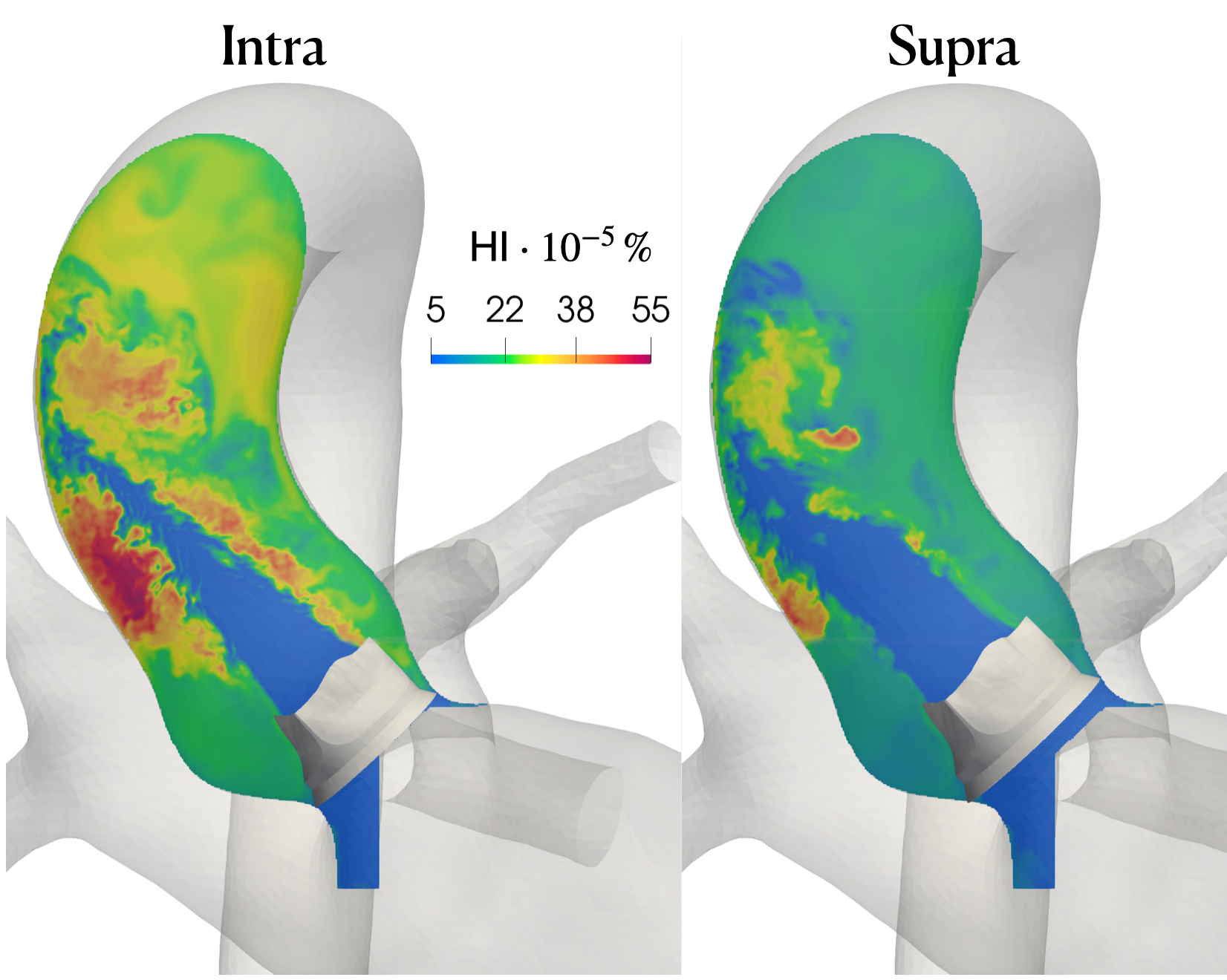}
  \caption{Eulerian stress-based hemolysis index $\HI(\pos,t)$ produced by aortic jet during systolic peak.}
  \label{fig:Figure_HI_systole}
\end{figure}


\section{Discussion}\label{sec:disc}
In this study, we employed a multyphysics computational model to evaluate how the deployment position of a prosthetic valve influences the cardiac hemodynamics. Specifically, we considered a realistic left heart model, including pulmonary veins, atrium, mitral valve, ventircle and thoracic aorta. By simulating physiological heart rate and stroke volume in a patient with a small aortic annulus, we replicated a typical surgical scenario where a small-sized heart requires prosthetic valve implantation. In such cases, surgeons must cope with several challenges that impact postoperative hemodynamic performance, with prosthesis-patient mismatch being a common concern. To investigate this, we analyzed the post-operative hemodynamics performance of a bioprosthetic valve resembling the LivaNova Crown PRT, implanted in both intra-annular and supra-annular configurations. This valve model was specifically chosen because, unlike most prosthetic valves, it can be implanted in either configuration. This unique feature allowed us to perform a controlled and high-fidelity numerical comparison, isolating the effects of deployment strategy on hemodynamics. In clinical practice, such a comparison is not possible since surgical implantation is irreversible. However, our computational framework enables this type of study, providing a rare opportunity to assess deployment choices under identical anatomical conditions. In addition to standard clinical metrics, we measured detailed hemodynamic quantities that are typically challenging to assess in clinical practice, providing valuable insights into the flow dynamics and mechanical environment associated with different deployment strategies.\\
\indent The results highlight the significant impact of valve size and positioning on hemodynamics, particularly in terms of ventricular workload, $\TPD$, valve dynamics, and blood damage metrics. One of the most relevant findings is the lower ventricular workload associated with the supra-annular configuration, which requires approximately 15~\unit{\mmHg} less ventricular pressure compared to the intra-annular position. This reduction suggests a more efficient ejection phase, likely driven by the increased $\GOA$. The supra-annular valve, indeed, exhibits a larger $\GOA$ (1.7 vs 1.45~\unit{\cm\squared} in the intra-annular case), which facilitates improved flow dynamics. Interestingly, while the absolute $\GOA$ is higher in the supra-annular configuration, the percentage $\GOA$ relative to the internal orifice diameter is slightly larger in the intra-annular case (53\% vs 51\%).\\
\indent Furthermore, $\EOA$ is characterized by slightly smaller values compared to $\GOA$, since it reflects the cross-sectional area where the highest flow velocities are reached. Consequently, the measured GOA values should not be comparad directly with the orifice area of clinical studies, where the EOA is reported. Clinical studies by \citet{takahashi2023_intra_supra_hemo_difference_MRI} report postoperative $\EOA$ values ranging from 1.2 to 1.8 \unit{\cm\squared}, which suggests that our computational setup represents a patient with a relatively small aortic annulus and body surface area. Indeed, the authors primarily analyzed patients receiving valves larger than 23~\unit{\mm}, indicative of a generally larger anatomical scale. Similar observations apply to studies by \citet{kim2019supra-intra-clinical-outcome-comparison} and \citet{okuno2019PPM_supra_intra}.\\
\indent Another key aspect revealed by the high spatial and temporal resolution of the computational model is the valve opening and closure dynamics. The simulations show that in both configurations, the valve takes approximately 120~\unit{\ms} to fully open and 80~\unit{\ms} to close. Additionally, the maximum $\GOA$ is not maintained throughout systole; instead, the valve reaches peak opening briefly before gradually reducing the orifice size, followed by a short plateau lasting approximately 100~\unit{\ms}. These measurements closely match the ones from natural valves, as investigated in \citet{handke2003aortic_valve_dynamics}. This dynamic behavior is often overlooked in clinical (imaging-based) lower-resolution studies but is essential for understanding flow patterns and potential valve leaflet stresses.\\
\indent $\TPD$ values further support the advantages of supra-annular positioning. Peak $\TPD$ reaches 30~\unit{\mmHg} in the intra-annular case and 23~\unit{\mmHg} in the supra-annular case, while mean values are 17~\unit{\mmHg} and 12~\unit{\mmHg}, respectively. These findings align well with previous clinical and in vitro studies, reinforcing the notion that supra-annular positioning reduces pressure gradients, improving overall hemodynamic efficiency. \citet{takahashi2023_intra_supra_hemo_difference_MRI} report mean pressure gradients of 13~\unit{\mmHg} for intra-annular valves and 8~\unit{\mmHg} for supra-annular valves, which closely match the computational results. Similarly, \citet{okuno2019PPM_supra_intra} report values of 8~\unit{\mmHg} and 10.7~\unit{\mmHg}, further validating the study’s findings.\\
\indent The analysis of $\WSS$ in the ascending aorta reveals another critical distinction. Due to the impingement of the systolic jet, peak $\WSS$ reaches approximately 40~\unit{\pascal} in the intra-annular configuration, compared to 25~\unit{\pascal} in the supra-annular case. This indicates that intra-annular positioning imposes significantly higher mechanical stresses on the aortic wall, which could have long-term implications for endothelial function and vascular remodeling. \citet{takahashi2023_intra_supra_hemo_difference_MRI} report postoperative average $\WSS$ reductions from 6.8~\unit{\pascal} to 4.8~\unit{\pascal} in the supra-annular case (2.0~\unit{\pascal} decrease) and from 6.8~\unit{\pascal} to 6.0~\unit{\pascal} in the intra-annular case (0.8~\unit{\pascal} decrease). Their peak $\WSS$ values also decrease from 53~\unit{\pascal} to 42~\unit{\pascal} (supra-annular) and from 53~\unit{\pascal} to 51~\unit{\pascal} (intra-annular). Although these trends qualitatively match the present results, direct quantitative comparison should be approached with caution, as $\WSS$ values obtained from MRI-based measurements are known to exhibit high uncertainty \cite{petersson2012wss_from_MRI_accuracy}.\\
\indent Finally, hemolysis comparison provides a direct assessment of blood damage risks. Both stress- and strain-based hemolysis models predict substantially higher RBC stresses in the intra-annular configuration. The difference in average stress levels stabilizes around 0.2~\unit{\pascal} (strain-based model) and 0.4~\unit{\pascal} (stress-based model). This trend is further confirmed by the Eulerian hemolysis model, which highlights markedly higher fluid viscous shear stress ($\viscstress$) in the shear layer of the intra-annular configuration, where the aortic jet reaches substantially higher peak velocities. These findings support the superior hemodynamic performance of the supra-annular configuration, particularly in terms of reducing blood damage and possibly improving long-term post-operative outcome.\\

\subsection{Limitations and future perspectives}\label{sec:limits}
Despite the effectiveness of the present numerical framework in capturing key hemodynamic features of aortic valve function, several limitations should be acknowledged, and various future research directions could further improve the understanding of prosthetic valve performance.\\
\indent The structural model employed in this study is based on a mass-spring discrete formulation, which effectively reproduces valve and aortic dynamics in terms of the investigated hemodynamic metrics. However, more advanced continuum-based structural models -- such as those employing orthotropic hyperelastic material properties -- could provide a more realistic representation of the mechanical behavior of the valve leaflets and aortic walls. These models, which have been extensively reported in the literature \cite{verzicco_2022_EFM,fedele2023comprehensive_quarteroni_heart_model,viola_2023_digital_twins}, would allow for a more detailed assessment of tissue-level mechanical stresses and strains, particularly in the case of bioprosthetic valves. Such an approach could be useful for studying valve durability, tissue fatigue, and stress-induced calcification, which are crucial factors affecting the long-term performance of prosthetic valves.\\
\indent Blood rheology is quite complex and shows a non-Newtonian shear-thinning behaviour \cite{verzicco_2022_EFM}. However, the non-Newtonian features become dominant only in submillimetric vessels and the flow developing in larger structures can be modeled by a Newtonian constitutive relation with good accuracy \citet{deVita2016non-newtonian}. Given that our study focuses on the primary cardiac structures, we adopt this Newtonian approximation, which has been widely validated in similar contexts. While we do not expect significant differences in the comparative trends between the two valve configurations, incorporating non-Newtonian rheology models could introduce small refinements to the computed hydrodynamic stress values and hemolysis predictions.\\
\indent Many of the observed hemodynamic improvements--such as reduced pressure gradients, lower peak velocities, and diminished shear stresses--are likely driven by the larger effective orifice area. Therefore, disentangling the effects of valve positioning from those of increased orifice area, which is a key feature of the supra-annular configuration, would be another interesting investigation. However, the isolated impact of different position might be worthy of a dedicated study in the future. One possible approach would involve placing a valve of identical size in both intra- and supra-annular positions. In the supra-annular case, this would require an artificial support structure (e.g., a stent) to secure the smaller valve. While this setup could help isolating the positional effects, it would also introduce additional geometric and mechanical parameters that could influence the flow, and such differences would need to be carefully accounted for in the analysis.\\
\indent The latter question becomes particularly relevant when considering coronary perfusion. The present study focuses on the hemodynamics of the aortic valve and the ascending aorta, without accounting for the coronary arteries, which may be influenced by transvalvular flow. Including the coronary circulation in future computational models could help assess whether the supra-annular valve position affects coronary flow. This aspect is particularly relevant in patients with coronary artery disease (CAD) or altered coronary anatomy, where even subtle changes in the flow distribution could impact myocardial perfusion.\\
\indent The hemolysis level of the two valve deployments has been evaluated numerically using various models, namely Lagrangian/stress-based, Lagrangian/strain-based and Eulerian/stress-based. It should be noted that these models predict different hemolysis level, and the dispersion of the results among the models for the same configuration may be larger than the hemolysis differences observed between the intra- and the supra- case. Such discrepancies between model predictions and experimental data have been already reported in the literature \cite{yu2017review_hemolysis}. For example, \citet{hashimoto1989RBCHemolysysExperiments} used a Couette-type flow with sinusoidal shear rates between 300 and 1000 s$^{-1}$ and, even under these simple conditions, different models can predict hemolysis values that differ by up to three orders of magnitude. Simpler models, like the stress-based approach, show better agreement with experimental data for steady, homogeneous shear flows. However, their applicability is limited to specific flow conditions, and their performance may degrade in more complex, non-uniform, or turbulent flows. Strain-based models generally offer better performance across a wider range of flow conditions, though experimental validation in such complex scenarios remains sparse \cite{faghih2019hemolysisReview2}. In conclusion, current hemolysis models lack sufficient validation, and there is no clear consensus on the required model complexity to ensure reliable predictions across diverse conditions. Nevertheless, the hemolysis models have been used here in a comparative manner to analyse hemolysis differences between the intra- and supra- deployment location. Importantly, all the hemolysis models considered here indicate significant reduction of the hemolysis level in the supra case, thus strengthening the result.\\
\indent Another important aspect for future investigation is the residence time of blood within the aortic root and valve region. Blood flow stagnation can lead to thrombus formation and increased risk of leaflet thrombosis, particularly in cases of suboptimal valve function. Consequently, blood residence time has been studied numerically in different cardiovascular districts such as the left atrial appendage \cite{garcia2021brt_auricle}. \citet{azadani2017blood_stasis} have explored blood stasis effects in transcatheter aortic valve procedures, where the interaction between degenerate native leaflets (valve-in-valve procedures) and newly implanted valves alters flow patterns. Also in this case, intra-annular positioning outlined worse hemodynamic performace compared to the other deployment position. A detailed Eulerian and Lagrangian analysis of blood residence time could provide valuable insights into potential thrombogenic regions, helping to optimize valve design and implantation strategies.\\
\indent The present study focuses on a single anatomical configuration, but patient-specific variability plays a crucial role in valve performance. Future in silico trials based on a population of patients with different anatomical and functional properties could provide a more comprehensive understanding of valve behavior under diverse physiological conditions. Relevant parameters to explore include: anatomical variability (differences in aortic root geometry, annulus diameter, and leaflet morphology), mechanical properties (such as stiffness and material model of the aortic wall or valve leaflets), cardiac function (ventricular ejection fraction, stroke volume, and heart rate) and valve type and model (mechanical or biological).\\

\section*{ACKNOWLEDGMENTS}
This paper represents a follow-up to the contribution presented by R. Verzicco at ICTAM 2024 (25-30 August, Daegu, Korea). This project has received funding from the European Research Council (ERC) under the European Union’s Horizon Europe research and innovation program (Grant No. 101039657, CARDIOTRIALS to F.V.). Part of the work has been supported by the grant ``Fluid dynamics of the right heart for early detection of disease development", MUR 2022AJT27Y (P.I. R. Verzicco). This work has received partial funding from the project MUR-FARE R2045J8XAW CUPE83C22005500001 (P.I. Luca Biferale).

\bibliography{refs}


\end{document}